\documentclass[manuscript]{aastex631}
\usepackage{amsmath}
\usepackage{multirow}
\usepackage{txfonts}
\usepackage{color}
\usepackage{graphicx}
\usepackage{longtable}
\usepackage[figuresright]{rotating}
\usepackage{times}
\usepackage{threeparttable}
\usepackage{hyperref}
\graphicspath{{./}{figures/}}

\begin{document}

\title{Correlations between Event Rates of Short Gamma-Ray Bursts and Star Formation Rates with/without Time Delay}

\author[0009-0007-9880-0610]{ X. Y. Du }
\affiliation{College of Physics and Physical Engineering, Qufu Normal University, Qufu 273165, P. R. China }
\author[0000-0003-4859-682X]{Z. B. Zhang}
\altaffiliation{Corresponding author: Zhi-Bin Zhang\\ zbzhang@qfnu.edu.com}
\affiliation{College of Physics and Physical Engineering, Qufu Normal University, Qufu 273165, P. R. China }
\author{ W. C. Du }
\affiliation{Engineering Laboratory for Optoelectronic Technology and Advanced Manufacturing, School of Physics, Henan Normal University, Xinxiang, 453007, P. R. China}
\author{ G. A. Li }
\affiliation{College of Physics and Physical Engineering, Qufu Normal University, Qufu 273165, P. R. China }
\author{ Y. Liu }
\affiliation{College of Physics and Physical Engineering, Qufu Normal University, Qufu 273165, P. R. China }
\author{ H. C. Liu }
\altaffiliation{H. C. Liu\\ hcliu12@sina.com}
\affiliation{Engineering Laboratory for Optoelectronic Technology and Advanced Manufacturing, School of Physics, Henan Normal University, Xinxiang, 453007, P. R. China}


\begin{abstract}

In this paper, we systematically investigate the redshift and luminosity distributions as well as the event rates of short Gamma-Ray Bursts (SGRBs) detected by Swift, Fermi, Konus-wind satellites. It is found that the distributions of redshift and luminosity of Fermi and Konus-wind SGRBs are identical and they obviously differ from those of Swift/BAT SGRBs. The luminosity distributions of SGRBs detected by diverse detectors can be uniformly fitted by a smoothly broken power-law function. The median luminosity of Swift SGRBs is about one order of magnitude smaller than that of Fermi/GBM or Konus-wind SGRBs. We also compare the local event rates of Swift/BAT, Fermi/GBM and Konus-wind SGRBs and find that the local rate of Swift SGRBs is around two orders of magnitude larger than that of either Fermi or Konus-wind SGRBs, while the latter two rates are  comparable. The observed SGRB rates can be successfully fitted by a power-law plus Gauss function. The SGRB rates of three kinds of detectors matche the delayed/undelayed SFRs well except the delayed Lognormal and/or Gaussian SFRs at higher redshift and exceed all types of SFRs at lower redshit of $z<1$ . After deducting the diverse SFR components from the SGRB rates, we surprisingly notice that the remaining SGRB rates steeply decline with redshift in a power-law-like form, indicating that these SGRBs could emerge from the old star populations or compact binary star mergers. 

\end{abstract}


\section{Introduction} \label{sec:intro}

Gamma-ray bursts (GRBs) are high-energy phenomena whose gamma rays suddenly intensify at a cosmological distance \citep{1986ApJ...308L..43P}. \cite{1993ApJ...413L.101K} statistically found that the durations  of prompt $\gamma$-ray emissions are bimodally distributed with a boundary of about two seconds \citep[see also][]{ 2008A&A...484..293Z}, namely long GRBs (LGRBs, $\mathrm{T_{90}} \textgreater$ 2s) originated from core-collapse of massive stars and short GRBs (SGRBs, $\mathrm{T_{90}} \textless$ 2s) produced by the merger of compact binary objects \citep{2018pgrb.book.....Z}. Thanks to the improvement of detection techniques, more and more telescopes provide us a wide range of observational data, so that the origin of different types of GRBs gets better understanding than before.

 One of interesting applications of GRBs in cosmology is that they can be used to probe the early history of star formation in the host galaxies. Particularly, luminosity function and event rate as two important physical quantities are of great significance to constrain the progenitors of diverse types of GRBs. Although the event rate of LGRBs has been comprehensively studied, the event rates of different samples of LGRBs estimated previously are inconsistent and their relationship with SFR remains controversal. Even for the same sample, the reduced event rates in diverse ways may be largely different. For example, some previous studies suggested that the LGRB event rates evolved along with the SFR \citep[e.g.][]{1997ApJ...486L..71T,1998MNRAS.294L..13W,2001ApJ...548..522P,2004RvMP...76.1143P,2004IJMPA..19.2385Z}. Other studies showed that the event rate of LGRBs exceeds the SFR at high redshift and while matches the SFR at low redshift \citep{2007ApJ...661..394L,2009ApJ...705L.104K,2022ApJ...938..129L}. On the contrary, some studies indicated that the event rate of LGRBs at low redshift exceeds the SFR  \citep{2015ApJ...806...44P,2015ApJS..218...13Y,2022MNRAS.513.1078D,2024MNRAS.527.7111L,2024ApJ...967L..30D}, which is mainly caused by the low-luminosity GRBs that usually occur at lower redshift  \citep{2023ApJ...958...37D,2024ApJ...963L..12P}.

 However, the increasing number of special GRBs has made the relation between GRB rate and SFR more complicated. For instance, GRB 060614 as a LGRB was found to lack an associated SN Ib/c \citep{2006Natur.444.1044G} and can also be treated as a short GRB with extended emission (EE) \citep{2007ApJ...655L..25Z,2015NatCo...6.7323Y}. The short GRB 200826A with $T_{90}=1.14$ s is instead produced from the core-collapse of massive star \citep{2021NatAs...5..911Z,2021MNRAS.503.2966R,2022ApJ...932....1R}. Interestingly,  GRB 211211A as a LGRB associated with \emph{kilonova} is found for the first time to originate from the WD-NS merger \citep{2022Natur.612..232Y,2022Natur.612..223R}.

Based on the extensive multiple-wavelength observations \citep{2017ApJ...848L..13A,2021ARA&A..59..155M}, it was confirmed that SGRB 170817A/GW170817  is formed from a binary neutron star merger, and is accompanied by a \emph{kilonova AT2017gfo} \citep{2017PhRvL.119p1101A}. Unlike the collapsing massive stars, binary stars must undergo a process of winding and coalescence which inevitably leads to visible time delay between the SGRB and the SFR. In other words, SGRBs could follow the delayed star formation \citep{2021MNRAS.501..157Z} \citep[e.g.][]{1998ApJ...498..106M,2006ApJ...650..281N,2011ApJ...727..109V,2012PhRvD..86b3502T,2018MNRAS.477.4275P,2021ApJ...917...24Z}. This may cause that the event rate of SGRBs is obviously different from the normal SFR \citep{2018ApJ...852....1Z}. Of course, some LGRBs with both lower luminosities \citep{2023ApJ...958...37D} and smaller redshifts \citep{2024ApJ...963L..12P} might also show a delay rate compared with the SFR if they are dominantly generated from the compact star mergers instead of the core-collapse. It is worth noting that those LGRBs with higher luminosities but smaller redshifts match the SFR well \citep[e.g.][]{2023ApJ...958...37D}. Even though the small number of SGRBs with measured redshift only occupy 10 percent of the observed GRB population \citep{2022ApJ...940....5D}, which may result in larger uncertainties of the reduced GRB rates. In addition, some instrumental and selection effects will also augment the fluctuation of the resulting event rate. Therefore, it is necessary to recalculate the event rates of SGRBs of larger samples under condition that the time delay is considered appropriately.

In the past several decades, three methods such as the non-parametric \citep[or Lynden-Bell $c^-$ method,][]{1971MNRAS.155...95L,1992ApJ...399..345E,2002ApJ...574..554L,2015ApJ...806...44P, 2021ApJ...914L..40D, 2022MNRAS.513.1078D, 2023ApJ...958...37D}, the parametric method \citep{2006ApJ...650..281N,2011ApJ...727..109V,2018MNRAS.477.4275P} and the maximum likelihood estimation \citep[MLE,][]{2019MNRAS.488.4607L,2024arXiv241117244H} had been frequently used to diagnose the relationship between the event rate of GRBs and the SFR. Excitingly, the inferred local event rate of SGRBs by means of different methods ranges from $0.1 \;  \mathrm{Gpc^{-3}yr^{-1}}$ to $400 \; \mathrm{Gpc^{-3}yr^{-1}}$ \citep{2006ApJ...650..281N,2011ApJ...727..109V,2015ApJ...812...33S,2018ApJ...852....1Z,2018MNRAS.477.4275P}, which may imply that the above three methods are somewhat self-consistent in a certain sense. In contrast, the advantage of parametric methods is that it can directly deal with the raw data without redshift deduction from the luminosity or energy as the non-parametric method did. Consequently, we will adopt the traditionally parametric method to investigate whether the SFR is indeed related with the rates of diverse kinds of SGRBs detected by different satellites.

The article is arranged as follows. Sample selection and data analysis are presented in Section \ref{sec:two}. The detailed methods are introduced in Section \ref{sec:three}. Our main results are shown in Section \ref{sec:four}. We will end with a summary in Section \ref{sec:five}.  The cosmological constants throughout the paper are taken as $\Omega_{\mathrm{M}}=0.3, \; \Omega_{\Lambda}=0.7, \; $and $\mathrm{H_{0}}$=70 $\mathrm{km \;s^{-1} \;Mpc^{-3}}$.

\section{Data preparation} \label{sec:two}
Fristly, we collected the SGRBs with measured redshift from the literature \citep{2023ApJ...950...30Z,2023ApJ...951....4G,2022ApJ...940...56F,2018PASP..130e4202Z,2022A&A...657A.124L}. As a result, 104 SGRBs are collected to constitute our sample in (Table \ref{tab:GRB}) , among which GRB 060614, GRB 100816A, GRB 211211A, GRB 211227A are merger-driven LGRBs \citep{2006Natur.444.1050D,2007ApJ...655L..25Z,2021NatAs...5..917A,2022Natur.612..223R,2022Natur.612..228T,2024ApJ...970....6X,2022ApJ...931L..23L,2023A&A...678A.142F}; GRB 200826A is collapse-driven SGRBs \citep{2021NatAs...5..911Z,2021NatAs...5..917A,2022ApJ...932....1R}; GRB 170817A is the first binary neutron star merger SGRBs and an off-axis burst \citep{2017Natur.551...85A}, and GRB 050709 detected only by HETE-II \citep{2005Natur.437..845F}, GRB 060614, GRB 061201, GRB 130603B, GRB 140903A and GRB 200522A are KN-associated SGRBs \citep{2024MNRAS.527.7111L}. It should be noted that these SGRBs were observed by distinct detectors with diverse performance parameters as shown in Table \ref{tab:Detectors}. The operation time (T), field of view ($\mathrm{\Omega}$) and energy flux sensitivity ($\mathrm{F_{th}}$) of detectors are three important parameters for the calculation of GRB rates. 

To reduce the instrumental effect on the calculation of GRB event rate, one should select the SGRBs recorded by a sole detector. Furthermore, the number of SGRBs in each sub-samples should be sufficient enough to ensure the statistical reliability. Eventually, we have chosen 103 SGRBs in total, including 95 Swift, 41 Fermi and 32 Konus-Wind SGRBs, of which 49, 43 and 11 out of them were simultaneously detected by one, two and three satellites, respectively.  

\begin{longrotatetable}
	\setlength{\tabcolsep}{3pt}
	\begin{deluxetable*}{ccccccccccccccc}
		\tablecaption{Characteristic parameters of SGRBs\label{tab:GRB}}
		\tabletypesize{\scriptsize}
		\tablehead{
			\colhead{GRB} & \colhead{T$_{90}$$^{f}$ (s)} & \colhead{z} & 
			\colhead{$\alpha  $} & \colhead{$\beta $} & 
			\colhead{P$^{d}$ ($\rm ph \;\rm cm^{-2}\;\rm s^{-1}$)} & 
			\colhead{S$^{f}$ ($\rm erg \; \rm cm^{-2}$)} &\colhead{E$_P$(keV)}  &\multicolumn{2}{c}{Range (keV)} &  \colhead{$P_{bolo}$$^{e}$ ($\rm erg \; \rm cm^{-2}\; \rm s^{-1}$)}& 
			\colhead{L ($\rm erg\; \rm s^{-1}$)}   &\colhead{$K$} & \colhead{Detectors}& \colhead{Ref$^{g}$}
		}
		\startdata
		050509B$\star$ & 0.073&0.2248&2.47&0&2.77$\pm$0.79&(5.85$\pm$1.62)$\times10^{-9}$&58.07$\pm$40.58&15&150&(2.13$\pm$0.61)$\times10^{-7}$&(3.19$\pm$0.91)$\times10^{49}$&1.02&Swift/BAT& (1,4.5,6,7,9)   \\\hline
		050709$^{a}$&0.07&0.1607&-0.70&0&12.10$\pm$0.40&(3.90$\pm$2.70)E-07&83$\pm$18&30&400& & & &HETE-II& (10) \\\hline
		050724$\star$&96&0.258&-1.58&0&3.35$\pm$0.30&(2.38$\pm$0.24)$\times10^{-7}$& 244.55 & 15 & 150 & (6.24$\pm$0.56)$\times10^{-7}$&(1.28$\pm$0.12)$\times10^{50}$& 2.64  & Swift/BAT  & (1,4,5,6,7,9)   \\\hline
		050813$\star$ & 0.45  & 1.8 & 1.28 & 0  & 0.45$\pm$0.17 & (3.47$\pm$1.39)$\times10^{-8}$ & 63.58$\pm$14.77  & 15 & 150 & (3.68$\pm$1.36)$\times10^{-8}$ & (8.22$\pm$3.04)$\times10^{50}$ & 1.05  & Swift/BAT  & (1,4,5,6,7,9)   \\\hline
		051210$\star$ & 1.3   & 2.58   & -0.57   & 0   & 0.76$\pm$0.13 & (7.35$\pm$1.23)$\times10^{-8}$ & 250.85$\pm$250.44 & 15 & 150  & (1.99$\pm$0.33)$\times10^{-7}$ & (1.07$\pm$0.18)$\times10^{52}$ & 2.70   & Swift/BAT  & (1,6,7,8,9)   \\\hline
		\multirow{2}{*}{051221A$\star$} & \multirow{2}{*}{1.4} & \multirow{2}{*}{0.5464}  & -1.08  & 0    & (4.60$\pm$0.20)$\times10^{-5}$ & (3.20$\pm$0.10)$\times10^{-6}$  & 402$\pm$93  & 20   & 2000  & (4.93$\pm$0.21)$\times10^{-5}$   & (5.86$\pm$0.23)$\times10^{52}$ & 1.07  & Konus-Wind & \multirow{2}{*}{(1,3,4,5,6,7,8,9)}   \\
		&    &   & -1.20    & 0     & 12.06$\pm$0.40 & (1.01$\pm$0.02)$\times10^{-6}$  & 657.55$\pm$657.77 & 15   & 150   & (4.79$\pm$0.16)$\times10^{-6}$ & (5.70$\pm$0.19)$\times10^{51}$ & 4.72  & Swift/BAT  &    \\\hline
		051227  & 114.6 & 0.8   & -0.99   & 0    & 0.95$\pm$0.15  & (8.95$\pm$1.57)$\times10^{-8}$  & 59.78$\pm$36.73$^{c}$ & 15   & 150   & (8.87$\pm$1.38)$\times10^{-8}$ & (2.68$\pm$0.42)$\times10^{50}$ & 1.42  & Swift/BAT  & (1,9)   \\\hline
		060121  & 2     & 0.154   & -0.51    & -2.39  & (1.64$\pm$0.18)$\times10^{-5}$ & (4.71$\pm$0.44)$\times10^{-6}$ & 134     $\pm$ 32       & 20   & 1000  & (2.83$\pm$0.31)$\times10^{-5}$ & (1.82$\pm$0.20)$\times10^{51}$ & 1.72  & Konus-Wind & (3,7,9)   \\\hline
		\multirow{2}{*}{060313$\star$}  & \multirow{2}{*}{0.74}  & \multirow{2}{*}{1.7}      & -0.60      & 0         & (7.00$\pm$1.00)$\times10^{-5}$ & (1.42$\pm$0.10)$\times10^{-5}$   & 922$\pm$ 306   & 20   & 2000  & (7.71 $\pm$ 1.10)$\times10^{-5}$ & (1.50$\pm$ 0.21)$\times10^{54}$ & 1.10   & Konus-Wind & \multirow{2}{*}{(1,3,6,7)}  \\
		&   &      & -0.68   & 0         & 10.85   $\pm$ 0.39     & (1.15$\pm$ 0.05)$\times10^{-6}$ & 122.18  $\pm$ 50.78 $^{c}$    & 15   & 150   & (1.43$\pm$0.05)$\times10^{-6}$ & (2.78$\pm$0.10)$\times10^{52}$ & 1.58  & Swift/BAT  &    \\\hline
		060502B$\star$ & 0.131 & 0.287    & -0.40   & 0         & 0.60$\pm$0.13 & (6.09$\pm$1.32)$\times10^{-8}$   & 231.43$\pm$230.76  & 15   & 150   & (1.57$\pm$0.33)$\times10^{-7}$ & (4.09$\pm$0.86)$\times10^{49}$ & 2.57  & Swift/BAT  & (1,4,5,9)   \\\hline
		\multirow{2}{*}{060614$\ast$}  & \multirow{2}{*}{108.7} & \multirow{2}{*}{0.125}    & -1.57       & 0         & (4.50$\pm$0.72)$\times10^{-6}$ & (8.19$\pm$ 0.56)$\times10^{-6}$  & 302$\pm$85   & 20   & 2000  & (6.02$\pm$0.96)$\times10^{-6}$ & (2.47$\pm$0.39)$\times10^{50}$ & 1.34  & Konus-Wind & \multirow{2}{*} {(1,3,5,6,7,9)}  \\
		&  &     & -1.38    & 0         & 11.39$\pm$0.73     & (8.28$\pm$0.57)$\times10^{-7}$   & 147.91$\pm$67.12  & 15   & 150   & (1.69$\pm$0.11)$\times10^{-6}$ & (6.92$\pm$0.44)$\times10^{49}$ & 2.04  & Swift/BAT  &    \\\hline
		060801$\star$  & 0.49  & 1.131    & -0.08  & 0  & 0.75$\pm$0.12 & (8.45$\pm$1.34)$\times10^{-8}$ & 272.78$\pm$272.76  & 15   & 150   & (2.80$\pm$0.44)$\times10^{-7}$ & (1.99$\pm$0.31)$\times10^{51}$ & 3.32  & Swift/BAT  & (1,6,7,8,9)   \\\hline
		\multirow{2}{*}{061006$\star$}  & \multirow{2}{*}{129.9} & \multirow{2}{*}{0.4377}   & -0.62       & 0         & (2.13$\pm$ 0.41)$\times10^{-5}$ & (3.57$\pm$ 0.31)$\times10^{-6}$   & 664     $\pm$ 227      & 20   & 2000  & (2.22$\pm$0.43)$\times10^{-5}$ & (1.56$\pm$ 0.30)$\times10^{52}$ & 1.04  & Konus-Wind & \multirow{2}{*}{(1,3,4,6,7,9)}   \\
		&  &    & -0.85   & 0         & 5.27    $\pm$ 0.21     & (5.24$\pm$0.19)$\times10^{-7}$  & 98.04$\pm$39.04 $^{c}$       & 15   & 150   & (5.99$\pm$0.24)$\times10^{-7}$ & (4.19$\pm$0.17)$\times10^{50}$ & 1.48  & Swift/BAT  &   \\\hline
		\multirow{2}{*}{061201$\star$}  & \multirow{2}{*}{0.76}  & \multirow{2}{*}{0.111}    & -0.36       & 0         & (3.19$\pm$0.72)$\times10^{-5}$ & (5.33$\pm$0.70)$\times10^{-6}$   & 873$\pm$ 458      & 20   & 3000  & (3.24$\pm$0.73)$\times10^{-5}$ & (1.03$\pm$0.23)$\times10^{51}$ & 1.02  & Konus-Wind & \multirow{2}{*}{(1,3,4,6,7,8,9)}   \\
		&    &      & -0.83   & 0         & 3.55    $\pm$ 0.35     & (3.56$\pm$0.37)$\times10^{-7}$   & 87.96$\pm$46.82 $^{c}$  & 15   & 150   & (3.78$\pm$0.38)$\times10^{-7}$ & (1.20$\pm$0.12)$\times10^{49}$ & 1.42  & Swift/BAT  &   \\\hline
		061210$\star$  & 85.3  & 0.4095   & -0.48   & 0   & 2.78$\pm$0.30 & (2.89$\pm$0.31)$\times10^{-7}$   & 346.19$\pm$346.18  & 15   & 150   & (1.12$\pm$0.12)$\times10^{-6}$  & (6.72$\pm$0.73)$\times10^{50}$ & 3.90   & Swift/BAT  & (1,6,7,9)  \\\hline
		061217$\star$  & 0.21  & 0.827    & -0.27   & 0         & 0.52$\pm$0.12     & (4.57$\pm$1.26)$\times10^{-8}$ & 109.85$\pm$         & 15   & 150   & (6.46$\pm$ 1.50)$\times10^{-8}$ & (2.11$\pm$0.49)$\times10^{50}$ & 1.41  & Swift/BAT  & (1,4,5,9)   \\\hline
		070429B & 0.47  & 0.902    & -0.37   & 0         & 1.06$\pm$0.16     & (6.23$\pm$ 1.27)$\times10^{-8}$  & 43.58$\pm$8.99  & 15   & 150   & (7.81$\pm$1.17)$\times10^{-8}$  & (3.17$\pm$0.48)$\times10^{50}$ & 1.25  & Swift/BAT  & (1,4,6,7,9)   \\\hline
		070714B$\star$ & 64    & 0.923    & -0.88    & 0         & 2.75$\pm$0.16     & (2.72$\pm$0.16)$\times10^{-7}$   & 81.62$\pm$36.76 $^{c}$   & 15   & 150   & (2.84$\pm$0.16)$\times10^{-7}$ & (1.22$\pm$0.07)$\times10^{51}$ & 1.42  & Swift/BAT  & (1,6,7,9)   \\\hline
		070724A$\star$ & 0.4   & 0.457    & 0.41     & 0         & 0.52$\pm$0.11     & (2.85$\pm$0.73)$\times10^{-8}$   & 38.77$\pm$8.95  & 15   & 150   & (3.36$\pm$0.7)0$\times10^{-8}$& (2.60$\pm$0.55)$\times10^{49}$ & 1.18  & Swift/BAT  & (1,4,6,7,8,9)  \\\hline
		070729  & 0.9   & 0.52     & -0.21   & 0         & 0.94$\pm$0.16     & (9.52$\pm$1.66)$\times10^{-8}$ & 181.86$\pm$181.88  & 15   & 150   & (1.97$\pm$0.34)$\times10^{-7}$ & (2.08$\pm$0.36)$\times10^{50}$ & 2.06  & Swift/BAT  & (1,6,7,9)   \\\hline
		070809$\star$  & 1.3   & 0.2187   & -1.33    & 0         & 1.21$\pm$0.15     & (8.88$\pm$1.38)$\times10^{-8}$   & 145.48$\pm$145.93  & 15   & 150   & (1.77$\pm$0.22)$\times10^{-7}$ & (2.49$\pm$ 0.30)$\times10^{49}$ & 1.99  & Swift/BAT  & (1,4,5,6,7,9)   \\\hline
		\multirow{2}{*}{071227$\star$}  & \multirow{2}{*}{1.8}   & \multirow{2}{*}{0.381} & -0.70  & 0  & (3.50$\pm$1.10)$\times10^{-6}$ & (1.60$\pm$0.20)$\times10^{-6}$   & 1000  & 20 & 1300  & (4.86$\pm$1.53)$\times10^{-6}$ & (2.45$\pm$ 0.77)$\times10^{51}$ & 1.39  & Konus-Wind & \multirow{2}{*}{(1,3,4,6,7,9)}   \\
		&     &      & -0.99   & 0         & 1.60$\pm$0.21  & (1.52$\pm$ 0.21)$\times10^{-7}$   & 69.31 $\pm$40.01 $^{c}$     & 15   & 150   & (1.56$\pm$0.21)$\times10^{-7}$ & (7.89$\pm$1.04)$\times10^{49}$ & 1.43  & Swift/BAT  &    \\\hline
		080123$\star$  & 115   & 0.495    & -1.13   & 0    & 1.43$\pm$0.40  & (1.29$\pm$ 0.35)$\times10^{-7}$  & 68.72& 15   & 150   & (1.45$\pm$0.41)$\times10^{-7}$ & (1.36$\pm$0.38)$\times10^{50}$ & 1.51  & Swift/BAT  & (1,6,7,9)  \\\hline
		080503$\star$  & 170   & 3.10 $^{b}$    & -1.23    & 0   & 0.67$\pm$0.13     & (5.26$\pm$1.17)$\times10^{-8}$  & 226.61  & 15   & 150   & (1.26$\pm$0.24)$\times10^{-7}$ & (1.06$\pm$ 0.20)$\times10^{52}$ & 2.40   & Swift/BAT  & (1,6,7)  \\\hline
		\multirow{2}{*}{080905A$\star$} & \multirow{2}{*}{1 }    & \multirow{2}{*}{0.1218}   & 0.19   & -2.37  & 6.32    $\pm$ 0.68     & (8.50$\pm$ 0.46)$\times10^{-7}$   & 317.18$\pm$52.54 & 10   & 1000  & (3.00$\pm$0.32)$\times10^{-6}$  & (1.16$\pm$ 0.12)$\times10^{50}$ & 1.74  & Fermi      & 	\multirow{2}{*}{(1,2,4,6,7,8,9)}  \\
		&       &     & -0.64   & 0   & 1.23$\pm$ 0.19     & (1.28$\pm$ 0.19)$\times10^{-7}$  & 870.91$\pm$880.93  & 15   & 150   & (1.34$\pm$ 0.20)$\times10^{-6}$ & (5.19$\pm$ 0.79)$\times10^{49}$ & 10.50  & Swift/BAT  &    \\\hline
		081024A & 1.8   & 3.05     & -1.07    & 0         & 1.09$\pm$ 0.13     & (2.24$\pm$ 0.00)$\times10^{-8}$   & 40.55  $^{c}$        & 15   & 150   & (9.60$\pm$1.11)$\times10^{-8}$ & (7.77 $\pm$ 0.89)$\times10^{51}$ & 1.53  & Swift/BAT  & (1,5,9)   \\\hline
		\multirow{2}{*}{081226A} & \multirow{2}{*}{0.4}   & \multirow{2}{*}{0.436 $^{b}$}    & -0.81  & -18.62 & 6.23    $\pm$ 1.37     & (4.30$\pm$ 0.23)$\times10^{-7}$   & 584.59$\pm$ 284.04 & 10   & 1000  & (1.98$\pm$   0.44)$\times10^{-6}$ & (1.37$\pm$ 0.30)$\times10^{51}$ & 1.22  & Fermi      & \multirow{2}{*}{(1,2,5,6,7)}   \\
		&     &     & -1.02     & 0         & 1.08$\pm$0.27     & (1.00$\pm$0.22)$\times10^{-7}$   & 61.72$\pm$  40.39 $^{c}$         & 15   & 150   & (1.02$\pm$0.26)$\times10^{-7}$ & (7.09$\pm$ 1.80)$\times10^{49}$ & 1.43  & Swift/BAT  &   \\\hline
		\multirow{2}{*}{090227B} & \multirow{2}{*}{0.304} & \multirow{2}{*}{1.61}     & -0.51  & -3.47 & 34.60$\pm$0.30      & (1.11$\pm$ 0.01)$\times10^{-5}$   & 2095.19$\pm$ 99.83 & 10   & 1000  & (4.90$\pm$0.01)$\times10^{-5}$   & (8.33$\pm$ 0.01)$\times10^{53}$ & 2.85  & Fermi      & \multirow{2}{*}{(2,3,9)}   \\
		&   &       & -0.43       & 0         & (3.98$\pm$ 0.58)$\times10^{-4}$ & (2.75$\pm$ 0.27)$\times10^{-5}$   & 2253$\pm$262      & 20   & 10000 & (3.35$\pm$  0.49)$\times10^{-4}$ & (5.69$\pm$ 0.83)$\times10^{54}$ & 0.84 & Konus-Wind &    \\\hline
		090305A & 0.4   & 1.147 $^{b}$    & -0.26   & 0         & 0.87$\pm$0.22     & (8.30$\pm$ 1.98)$\times10^{-8}$   & 145.63$\pm$145.62  & 15   & 150   & (1.41$\pm$0.37)$\times10^{-7}$ &( 1.04 $\pm$ 0.27)$\times10^{51}$ & 1.70   & Swift/BAT  & (1,6,7)   \\\hline
		090426$\star$  & 1.2   & 2.609    & -1.16    & 0         & 2.25    $\pm$ 0.31     & (1.50$\pm$ 0.25)$\times10^{-7}$   & 67.22$\pm$22.29  & 15   & 150   & (2.32$\pm$0.32)$\times10^{-7}$  & (1.28$\pm$ 0.18)$\times10^{52}$ & 1.54  & Swift/BAT  & (1,4,5,6,7,8,9)   \\\hline
		\multirow{3}{*}{090510$\star$}  & \multirow{3}{*}{0.3}   & \multirow{3}{*}{0.903}    & -0.84  & -2.57 & 40.95$\pm$1.58     & (3.37$\pm$ 0.04)$\times10^{-6}$   & 4248.14$\pm$ 440.05 & 10   & 1000  & (6.35$\pm$  0.25)$\times10^{-5}$  & (2.58$\pm$ 0.10)$\times10^{53}$ & 3.98  & Fermi      & \multirow{3}{*}{(1,2,3,4,5,6,7,8,9)}   \\
		&     &      & 0.11        & -1.61     & (2.27$\pm$ 0.63)$\times10^{-4}$ & (2.42$\pm$ 0.37)$\times10^{-5}$   & 552$\pm$ 249      & 20   & 15000 &  (6.10$\pm$  1.69)$\times10^{-4}$ & (2.48 $\pm$ 0.69)$\times10^{54}$ & 2.69  & Konus-Wind &    \\
		&     &      & -0.57   & 0         & 4.34    $\pm$ 0.55     & (4.26$\pm$ 0.55)$\times10^{-7}$   & 264.49 $\pm$ 264.48  & 15   & 150   & (1.22$\pm$ 0.15)$\times10^{-6}$ & (4.94$\pm$ 0.63)$\times10^{51}$ & 2.85  & Swift/BAT  &    \\\hline
		090515$\star$  & 0.036 & 0.403    & -0.33   & 0         & 0.29$\pm$0.13     & (2.66$\pm$ 1.35)$\times10^{-8}$  & 137.78$\pm$ 137.78  & 15   & 150   & (4.36$\pm$  1.98)$\times10^{-8}$ & (2.51 $\pm$ 1.14)$\times10^{49}$& 1.64  & Swift/BAT  & (1,6,7,9)  \\\hline
		\multirow{2}{*}{090927}  & \multirow{2}{*}{2.2}   & \multirow{2}{*}{1.37}     & -0.73  & -9.54 & 6.54$\pm$1.09     & (3.03$\pm$ 0.18)$\times10^{-7}$   & 195.22$\pm$69.05 & 10   & 1000  & (9.12$\pm$  1.52)$\times10^{-7}$ & (1.04$\pm$ 0.17)$\times10^{52}$ & 1.03  & Fermi      & \multirow{2}{*}{(1,2,4,9)}   \\
		&     &       & -0.67   & 0         & 1.83    $\pm$ 0.22     & (1.18$\pm$ 0.20)$\times10^{-7}$  & 55.14 $\pm$ 12.03  & 15   & 150   & (1.53$\pm$0.19)$\times10^{-7}$ & (1.75$\pm$0.21)$\times10^{51}$ & 1.29  & Swift/BAT  &    \\\hline
		091109B & 0.3   & 0.147 $^{b}$    & -0.76   & 0         & 1.90     $\pm$ 0.18     & (6.48$\pm$ 0.00)$\times10^{-9}$  & 28.66 $^{c}$ & 15   & 150   & (1.47$\pm$ 0.14)$\times10^{-7}$ & (8.58$\pm$ 0.81)$\times10^{48}$ & 1.57  & Swift/BAT  & (1,5,6,7)  \\\hline
		\multirow{2}{*}{091117}  & \multirow{2}{*}{0.43}  & \multirow{2}{*}{0.096}    & -0.42       & -9.40      & (1.40$\pm$ 0.70)$\times10^{-5}$ & (1.90$\pm$ 0.50)$\times10^{-6}$   & 663$\pm$225      & 20   & 2000  & (1.45$\pm$  0.72)$\times10^{-5}$ &(3.37$\pm$1.68)$\times10^{50}$ & 1.03  & Konus-Wind & \multirow{2}{*}{(1,3,9)}   \\
		&    &      & -0.75   & 0         & 2.49    $\pm$ 0.58     & (4.13$\pm$ 0.00)$\times10^{-9}$   & 911.26   & 15   & 150   & (2.44$\pm$ 0.57)$\times10^{-6}$ & (5.67$\pm$ 1.33)$\times10^{49}$ & 9.74  & Swift/BAT  &    \\\hline
		\multirow{2}{*}{100117A$\star$} & \multirow{2}{*}{0.3}   & \multirow{2}{*}{0.914}    & -0.10  & -6.29 & 7.95$\pm$0.86     & (4.23$\pm$ 0.69)$\times10^{-7}$   & 327.22$\pm$ 52.92 & 10   & 1000  & (2.15$\pm$  0.23)$\times10^{-6}$  & (9.02$\pm$ 0.97)$\times10^{51}$ & 1.02  & Fermi      & \multirow{2}{*}{(1,2,5,6,7,8,9)}   \\
		&     &      & -0.55   & 0         & 2.90     $\pm$ 0.39     & (9.26$\pm$ 1.31)$\times10^{-8}$   & 265.76$\pm$265.76  & 15   & 150   & (8.22$\pm$1.12)$\times10^{-7}$ & (3.44$\pm$0.47)$\times10^{51}$ & 2.87  & Swift/BAT  &   \\\hline
		\multirow{2}{*}{100206A$\star$} & \multirow{2}{*}{0.12}  & \multirow{2}{*}{0.407}    & -0.32  & -2.26 & 25.37$\pm$ 1.17   & (7.57$\pm$ 0.11)$\times10^{-7}$& 454.31  $\pm$ 63.63 & 10   & 1000  & (1.34$\pm$ 0.06)$\times10^{-5}$   & (7.90$\pm$ 0.36)$\times10^{51}$ & 1.72  & Fermi      & \multirow{2}{*}{(1,2,5,6,7,9)}   \\
		&    &     & -0.78   & 0         & 1.42    $\pm$ 0.20      & (1.45$\pm$ 0.21)$\times10^{-7}$  & 68.47   $\pm$  40.03 $^{c}$        & 15   & 150   & (1.32$\pm$ 0.19)$\times10^{-7}$& (7.80$\pm$ 1.10)$\times10^{49}$& 1.34  & Swift/BAT  &    \\\hline
		\multirow{3}{*}{100625A$\star$} & \multirow{3}{*}{0.33}  & \multirow{3}{*}{0.452}    & -0.59  & -12.23 & 17.08 $\pm$ 2.62     & (5.63$\pm$0.25)$\times10^{-7}$   & 482.13$\pm$61.94  & 10   & 1000  & (5.27$\pm$0.81)$\times10^{-6}$  & (3.98$\pm$ 0.61)$\times10^{51}$ & 1.12  & Fermi      & \multirow{3}{*}{(1,2,3,5,6,7,8,9)}   \\
		&    &      & -0.10        & 0         &( 8.10$\pm$1.50)$\times10^{-6}$ & (8.30$\pm$ 1.50)$\times10^{-7}$   & 414     $\pm$ 78       & 20   & 2000  & (8.15$\pm$  1.51)$\times10^{-6}$ & (6.16 $\pm$ 1.14)$\times10^{51}$& 1.01  & Konus-Wind &    \\
		&    &      & -0.73   & 0         & 2.54    $\pm$ 0.18     & (2.46$\pm$ 0.19)$\times10^{-7}$  & 391.38$\pm$391.39  & 15   & 150   & (9.80$\pm$0.69)$\times10^{-7}$  & (7.40$\pm$0.52)$\times10^{50}$ & 3.98  & Swift/BAT  &    \\\hline
		100724A$\star$ & 1.40   & 1.288    & 0.04    & 0         & 1.80$\pm$0.24     & (1.07$\pm$ 0.17)$\times10^{-7}$& 45.15$\pm$ 7.09  & 15   & 150   & (1.26$\pm$0.17)$\times10^{-7}$    & (1.24 $\pm$ 0.16)$\times10^{51}$ & 1.18  & Swift/BAT  & (1,4,5,8)   \\\hline
		\multirow{3}{*}{100816A$\ast$} & \multirow{3}{*}{2.9}   & \multirow{3}{*}{0.8}      & -0.32  & -2.73 & 19.88$\pm$1.08     & (3.65$\pm$ 0.05)$\times10^{-6}$   & 133.13$\pm$7.08 & 10   & 1000  & (2.96$\pm$0.16)$\times10^{-6}$   & (8.94$\pm$ 0.49)$\times10^{51}$ & 1.34  & Fermi      & \multirow{3}{*}{(1,2,3,9)}   \\
		&     &        & -1.00          & 0         & (2.30$\pm$0.40)$\times10^{-6}$& (3.30$\pm$0.40)$\times10^{-6}$ & 148    $\pm$ 26       & 20   & 2000  & (2.62$\pm$0.46)$\times10^{-6}$ & (7.91 $\pm$ 1.38)$\times10^{51}$& 1.14  & Konus-Wind &    \\
		&     &       & -0.49   & 0         & 10.83   $\pm$ 0.45     & (1.61$\pm$ 0.28)$\times10^{-7}$   & 153.04$\pm$32.02  & 15   & 150   & (1.77$\pm$0.07)$\times10^{-6}$  & (5.33$\pm$0.22)$\times10^{51}$ & 1.78  & Swift/BAT  &    \\\hline
		\multirow{2}{*}{101219A$\star$} & \multirow{2}{*}{0.6}   & \multirow{2}{*}{0.7179}   & -0.22       & 0         & (2.80$\pm$ 0.80)$\times10^{-5}$ & (3.60$\pm$ 0.50)$\times10^{-6}$   & 490     $\pm$ 79       & 20   & 10000 & (2.82$\pm$0.80)$\times10^{-5}$ & (6.51$\pm$ 1.86)$\times10^{52}$ & 1.01  & Konus-Wind & \multirow{2}{*}{(1,3,4,5,6,7,8,9)}   \\
		&     &     & -0.59   & 0         & 3.87    $\pm$ 0.20      & (4.26$\pm$0.22)$\times10^{-7}$   & 92.54   $\pm$  40.31 $^{c}$        & 15   & 150   &( 4.19$\pm$0.22)$\times10^{-7}$ & (9.68 $\pm$ 0.51)$\times10^{50}$ & 1.37  & Swift/BAT  &  \\\hline
		\multirow{2}{*}{101224A} & \multirow{2}{*}{0.2}   & \multirow{2}{*}{0.454}    & -0.83   & -1.64 & 6.71    $\pm$ 1.04     & (1.91$\pm$0.27)$\times10^{-7}$  & 253.47  $\pm$ 217.99 & 10   & 1000  & (3.39$\pm$0.53)$\times10^{-6}$  & (2.59$\pm$ 0.40)$\times10^{51}$ & 3.46  & Fermi      & \multirow{2}{*}{(1,2,5,6,7,9)}   \\
		&     &      & -0.05  & 0         & 0.68    $\pm$ 0.19     & (7.09$\pm$ 2.02)$\times10^{-8}$  & 175.87  $\pm$ 176.03  & 15   & 150   & (1.43$\pm$  0.41)$\times10^{-7}$ & (1.09$\pm$0.31)$\times10^{50}$ & 2.01  & Swift/BAT  &   \\\hline
		110112A & 0.5   & 0.53 $^{b}$     & 0.09   & 0         & 0.48    $\pm$ 0.15     & (2.74$\pm$ 1.01)$\times10^{-8}$  & 42.27   & 15   & 150   & (3.27$\pm$1.02)$\times10^{-8}$& (3.62 $\pm$ 1.13)$\times10^{49}$ & 1.19  & Swift/BAT  & (1,6,7)   \\\hline
		\multirow{2}{*}{111117A$\star$} & \multirow{2}{*}{0.47}  & \multirow{2}{*}{2.211}    & -0.46  & -2.57 & 10.93   $\pm$ 1.06     &( 5.64$\pm$0.13)$\times10^{-7}$  & 502.85  $\pm$ 111.18 & 10   & 1000  & (4.45$\pm$0.43)$\times10^{-6}$  & (1.65$\pm$ 0.16)$\times10^{53}$ & 1.34  & Fermi      & \multirow{2}{*}{(1,2,4,5,6,7,8,9)}   \\
		&    &      & -0.27   & 0         & 1.35    $\pm$ 0.20      & (7.51$\pm$ 0.00)$\times10^{-10}$  & 350.39  $\pm$ 350.40    & 15   & 150   & (6.37$\pm$  0.97)$\times10^{-7}$ & (2.35$\pm$ 0.36)$\times10^{52}$ & 4.29  & Swift/BAT  &    \\\hline
		120305A$\star$ & 0.1   & 0.225    & -0.83   & 0         & 2.19    $\pm$ 0.16     & (2.02$\pm$ 0.17)$\times10^{-7}$   & 313.17  $\pm$ 313.18  & 15   & 150   &(6.28$\pm$0.45)$\times10^{-7}$ & (9.42$\pm$ 0.68)$\times10^{49}$ & 3.12  & Swift/BAT  & (1,5,6,7,8,9)  \\\hline
		\multirow{2}{*}{120804A$\star$} & \multirow{2}{*}{0.81}  & \multirow{2}{*}{1}        & -0.89       & 0         & (6.00$\pm$ 2.70)$\times10^{-6}$ &( 1.45$\pm$0.30)$\times10^{-6}$  & 135     $\pm$ 29       & 15   & 1000  & (6.55 $\pm$  2.95)$\times10^{-6}$& (3.43$\pm$ 1.54)$\times10^{52}$ & 1.09  & Konus-Wind & \multirow{2}{*}{(1,3,5,6,7,8,9)}   \\
		&    &      & -0.97   & 0         & 10.64   $\pm$ 0.57     & (8.66$\pm$ 0.47)$\times10^{-7}$   & 156.16  $\pm$ 50.42  & 15   & 150   & (1.64$\pm$  0.09)$\times10^{-6}$ & (8.56$\pm$0.46)$\times10^{51}$ & 1.89  & Swift/BAT  &    \\\hline
		121226A$\star$ & 1     & 1.37     & -1.48   & 0         & 0.83    $\pm$ 0.44     & (7.11$\pm$ 2.15)$\times10^{-8}$   & 56.05$\pm$  40.07 $^{c}$  & 15 & 150 & (9.51$\pm$ 5.06)$\times10^{-8}$& (1.08$\pm$ 0.58)$\times10^{51}$ & 1.83  & Swift/BAT  & (1,6,7,8)   \\\hline
		\multirow{3}{*}{130515A} & \multirow{3}{*}{0.29}  &\multirow{3}{*}{ 0.8}      & -0.26  & -2.63 & 20.60$\pm$ 2.33     & (1.09$\pm$ 0.02)$\times10^{-6}$  & 452.86$\pm$ 74.42 & 10   & 1000  & (9.17$\pm$1.04)$\times10^{-6}$& (2.77$\pm$0.31)$\times10^{52}$& 1.4   & Fermi      & \multirow{3}{*}{(1,2,3,5,6,7,9)}   \\
		&    &        & -0.50        & 0         & (2.10$\pm$0.50)$\times10^{-5}$ & (1.10$\pm$ 0.20)$\times10^{-6}$   & 715     $\pm$ 360      & 20   & 1500  & (2.34$\pm$  0.56)$\times10^{-5}$ & (7.07$\pm$ 1.68)$\times10^{52}$ & 1.12  & Konus-Wind &    \\
		&    &        & 0.18    & 0         & 1.37    $\pm$ 0.21     & (1.35$\pm$ 0.22)$\times10^{-7}$   & 122.45  $\pm$ 38.91  & 15   & 150   & (1.97$\pm$0.31)$\times10^{-7}$ &( 5.94$\pm$ 0.92)$\times10^{50}$ & 1.46  & Swift/BAT  &   \\\hline
		\multirow{2}{*}{130603B$\star$} & \multirow{2}{*}{0.18}  & \multirow{2}{*}{0.3568}   & -0.73       & 0         & (1.00$\pm$ 0.20)$\times10^{4}$ &( 6.60$\pm$0.70)$\times10^{-6}$   & 660     $\pm$ 100      & 20   & 10000 & (1.01$\pm$0.20)$\times10^{-4}$ & (4.38$\pm$ 0.88)$\times10^{52}$ & 1.01  & Konus-Wind & \multirow{2}{*}{(1,3,4,5,6,7,9)}   \\
		&    &     & -0.84  & 0         & 6.36    $\pm$ 0.25     & (6.34$\pm$ 0.28)$\times10^{-7}$  & 103.42  $\pm $  43.02 $^{c}$       & 15   & 150   & (7.46$\pm$0.29)$\times10^{-7}$ & (3.23$\pm$ 0.13)$\times10^{50}$ & 1.51  & Swift/BAT  &    \\\hline
		\multirow{2}{*}{130716A$\star$} & \multirow{2}{*}{0.77}  &\multirow{2}{*}{ 2.2}      & -0.51  & -4.93 & 5.73    $\pm$ 0.90      & (6.40$\pm$ 0.09)$\times10^{-7}$   & 883.47   $\pm$ 1120.96 & 10   & 1000  & (3.37$\pm$0.53)$\times10^{-6}$ & (1.23$\pm$ 0.19)$\times10^{53}$ & 1.49  & Fermi      & \multirow{2}{*}{(1,2,6,7,9)}   \\
		&    &        & -0.89  & 0         & 0.98    $\pm$ 0.20      & (9.39$\pm$ 0.00)$\times10^{-9}$  & 31.79   $^{c}$        & 15   & 150   & (8.04$\pm$  1.62)$\times10^{-8}$& (2.93$\pm$ 0.59)$\times10^{51}$ & 1.56  & Swift/BAT  &    \\\hline
		130822A & 0.044 & 0.154    & -0.99  & 0         & 0.31    $\pm$ 0.12     & (1.99$\pm$ 0.00)$\times10^{-8}$  & 53.24   & 15   & 150   &( 2.82$\pm$  1.09)$\times10^{-8}$& (1.82 $\pm$ 0.71)$\times10^{48}$ & 1.42  & Swift/BAT  & (1,6,7,9)   \\\hline
		\multirow{3}{*}{130912A} & \multirow{3}{*}{0.28}  & \multirow{3}{*}{0.617 $^{b}$}    & -0.81  & -1.66 & 15.56   $\pm$ 1.47     & (7.01$\pm$ 0.18)$\times10^{-7}$  & 407.68  $\pm$ 163.61 & 10   & 1000  & (9.02$\pm$  0.85)$\times10^{-6}$ & (1.44$\pm$ 0.14)$\times10^{52}$ & 2.79  & Fermi      & \multirow{3}{*}{( 1,2,3,5,6,7)}  \\
		&    &      & -1.07       & 0   & (2.20$\pm$ 0.60)$\times10^{-5}$ & (1.60$\pm$ 0.20)$\times10^{-6}$   & 1020     $\pm$ 320      & 20   & 10000 & (2.25$\pm$0.61)$\times10^{-5}$ &( 3.59$\pm$ 0.98)$\times10^{52}$ & 1.02  & Konus-Wind &    \\
		&    &      & -1.20    & 0         & 2.14    $\pm$ 0.34     & (1.79$\pm$ 0.29)$\times10^{-7}$  & 27.88  & 15   & 150   & (1.94$\pm$  0.31)$\times10^{-7}$ & (7.80$\pm$ 1.23)$\times10^{48}$ & 1.76  & Swift/BAT  &   \\\hline
		\multirow{2}{*}{131004A$\star$} & \multirow{2}{*}{1.54}  & \multirow{2}{*}{0.717}    & -1.36   & -1.66    & 9.82    $\pm$ 1.71     & (5.10$\pm$ 0.19)$\times10^{-7}$  & 118.13   $\pm$ 24.42 & 10   & 1000  & (2.78$\pm$0.49)$\times10^{-6}$ & (6.41$\pm$ 1.12)$\times10^{51}$& 3.85  & Fermi      & \multirow{2}{*}{(1,2,4,5,6,7,8,9)}   \\
		&    &      & -0.85   & 0         & 3.27    $\pm$ 0.21     & (2.21$\pm$0.17)$\times10^{-7}$   & 63.37   $\pm$ 9.49   & 15   & 150   & (2.99$\pm$0.19)$\times10^{-7}$ & (6.90$\pm$ 0.45)$\times10^{50}$& 1.36  & Swift/BAT  &    \\\hline
		140516A$\star$ & 0.19  & 0.618 $^{b}$    & 0.26    & 0         & 0.47    $\pm$ 0.15     & (2.95$\pm$ 1.02)$\times10^{-8}$  & 48.05 & 15   & 150   & (3.34$\pm$1.08)$\times10^{-8}$& (5.36$\pm$ 1.73)$\times10^{49}$& 1.14  & Swift/BAT  & (1,6,7,8)   \\\hline
		140619B & 0.5   & 2.67     & -0.17  & -2.87 & 5.28    $\pm$ 0.75     & (1.55$\pm$ 0.07)$\times10^{-6}$   & 1306.56 $\pm$ 297.46 & 10   & 1000  & (5.90$\pm$ 0.84)$\times10^{-6}$ & (3.45$\pm$ 0.49)$\times10^{53}$ & 1.99  & Fermi      & (2,4,9)   \\\hline
		140622A & 0.13  & 0.959    & -1.62    & 0         & 0.63    $\pm$ 0.23     & (3.88$\pm$ 1.86)$\times10^{-8}$  & 52.24   & 15   & 150   & (7.81$\pm$  2.84)$\times10^{-8}$ & (3.68 $\pm$ 1.34)E$\times10^{50}$& 2.01  & Swift/BAT  & (1,4,6,7,9)   \\\hline
		140903A$\star$ & 0.3   & 0.3529   & -1.56    & 0         & 2.45    $\pm$ 0.19     & 0.00   & 40.25   & 15   & 150   & (2.77$\pm$  0.22)$\times10^{-7}$ & (1.16 $\pm$ 0.09)$\times10^{50}$ & 1.90   & Swift/BAT  & (1,4,5,6,7)   \\\hline
		\multirow{2}{*}{140930B$\star$} & \multirow{2}{*}{0.84}  & \multirow{2}{*}{1.465}    & -0.60        & 0         & (3.40$\pm$ 1.10)$\times10^{-5}$ & (8.10$\pm$ 2.50)$\times10^{-6}$  & 1302    $\pm$ 459      & 20   & 10000 & (3.32$\pm$  1.07)$\times10^{-5}$ & (4.46$\pm$ 1.44)$\times10^{53}$ & 0.97 & Konus-Wind & \multirow{2}{*}{(1,3,5,7,8,9)}   \\
		&    &     & -0.64   & 0         & 3.99    $\pm$ 0.40      & (4.30$\pm$ 0.45)$\times10^{-7}$   & 92.77   $\pm $    49.19 $^{c}$     & 15   & 150   & (4.34$\pm $ 0.44)$\times10^{-7}$& (5.83 $\pm$ 0.59)$\times10^{51}$ & 1.39  & Swift/BAT  &    \\\hline
		141212A$\star$ & 0.3   & 0.596    & -0.48   & 0         & 1.19    $\pm$ 0.21     &( 8.71$\pm$ 1.81)$\times10^{-8}$   & 72.76   $\pm$ 20.80  & 15   & 150   & (1.10$\pm$0.19)$\times10^{-7}$ & (1.61$\pm$ 0.29)$\times10^{50}$ & 1.26  & Swift/BAT  & (1,4,5,6,7,9)   \\\hline
		\multirow{2}{*}{150101B$\star$} & \multirow{2}{*}{0.018} & \multirow{2}{*}{0.093}    & 1.65   & -1.94 & 10.48   $\pm$ 1.35     & (2.38$\pm$ 0.15)$\times10^{-7}$   & 28.67  $\pm$ 6.74 & 10   & 1000  & (1.76$\pm$  0.23)$\times10^{-6}$    & (3.82$\pm$ 0.49)$\times10^{49}$ & 6.07  & Fermi      & \multirow{2}{*}{(1,2,4,5,6,7,9)}   \\
		&   &     & -0.58   & 0         & 0.73    $\pm$ 0.30      & (5.85$\pm$ 2.33)$\times10^{-8}$ & 96.55    $\pm$ 96.45  & 15   & 150   & (8.16$\pm$3.32)$\times10^{-8}$& (1.77$\pm$0.72)$\times10^{48}$ & 1.39  & Swift/BAT  &    \\\hline
		\multirow{2}{*}{150120A$\star$} & \multirow{2}{*}{1.2}   & \multirow{2}{*}{0.4604}   & -1.43       & -1.65 & 3.10     $\pm$ 0.30      & (3.40$\pm$ 0.80)$\times10^{-7}$   & 130     $\pm$ 50       & 10   & 1000  & (9.36$\pm$0.91)$\times10^{-7}$ & (7.39$\pm$ 0.72)$\times10^{50}$ & 3.96  & Fermi      & \multirow{2}{*}{(1,2,4,5,6,7,8,9)}   \\
		&     &     & -1.79     & 0         & 1.77    $\pm$ 0.18     & (1.19$\pm$ 0.11)$\times10^{-7}$   & 64.69     $\pm$  33.58 $^{c}$        & 15   & 150   & (2.64$\pm$0.27)$\times10^{-7}$ & (2.08 $\pm$ 0.21)$\times10^{50}$ & 2.38  & Swift/BAT  &   \\\hline
		150423A$\star$ & 0.22  & 1.394    & 0.01  & 0         & 0.83    $\pm$ 0.15     & (8.17$\pm$ 1.63)$\times10^{-8}$  & 134.95  & 15   & 150   & (1.30 $\pm$  0.24)$\times10^{-7}$ & (1.54$\pm$0.28)$\times10^{51}$ & 1.59  & Swift/BAT  & (1,4,5,6,7,9)   \\\hline
		\multirow{2}{*}{150424A$\star$} & \multirow{2}{*}{91}    & \multirow{2}{*}{0.3}      & -0.37       & 0         & (1.85$\pm$0.48)$\times10^{-4}$ & (1.81$\pm$ 0.11)$\times10^{-5}$  & 919      $\pm$ 76       & 20   & 10000 & (1.86$\pm$0.48)$\times10^{-4}$ & (5.36$\pm$ 1.39)$\times10^{52}$ & 1.00     & Konus-Wind & \multirow{2}{*}{(1,3,6,7,9)}   \\
		&      &        & -0.79   & 0         & 0.07    $\pm$ 0.08     & 2.67$\pm$ 0.00$\times10^{-8}$  & 42.59      $^{c}$      & 15   & 150   &( 5.34$\pm$  6.65)$\times10^{-9}$ & (1.54 $\pm$ 1.92)$\times10^{48}$ & 1.38  & Swift/BAT  &    \\\hline
		150728A & 0.83  & 0.461    & -2.00          & 0         & 0.54    $\pm$ 0.12     & 0   & 72.35    & 15   & 150   & (1.33$\pm$  0.30)$\times10^{-7}$ &   (1.06 $\pm$ 0.24)$\times10^{50}$ & 4.00     & Swift/BAT  & (1,6,7,9)  \\\hline
		\multirow{2}{*}{150831A} & \multirow{2}{*}{1.15}  & \multirow{2}{*}{1.18}     & -0.50        & 0         & (9.10$\pm$ 2.50)$\times10^{-6}$& (2.40$\pm$ 0.40)$\times10^{-6}$  & 564     $\pm$ 122      & 20   & 10000 & (9.18$\pm$  2.52)$\times10^{-6}$ & (7.23$\pm$ 1.99)$\times10^{52}$ & 1.01  & Konus-Wind & \multirow{2}{*}{(1,3,5,6,7,9)}   \\
		&    &       & -0.70   & 0         & 3.35    $\pm$ 0.28     & (3.53$\pm$ 0.28)$\times10^{-7}$  & 87.78    $\pm$  43.28 $^{c}$         & 15   & 150   & (3.53$\pm$  0.29)$\times10^{-7}$ & (2.78 $\pm$ 0.23)$\times10^{51}$ & 1.38  & Swift/BAT  &    \\\hline
		\multirow{2}{*}{151229A$\star$} & \multirow{2}{*}{1.78}  & \multirow{2}{*}{1.4}      & -1.31   & -10.42 & 15.74   $\pm$ 1.55     & (1.11$\pm$ 0.02)$\times10^{-6}$  & 104.02   $\pm$ 14.15 & 10   & 1000  & (1.42$\pm$0.14)$\times10^{-6}$ & (1.70$\pm$0.17)$\times10^{52}$ & 1.17  & Fermi      & \multirow{2}{*}{(1,2,5,6,7,8,9)}   \\
		&   &       & -1.44    & 0         & 7.08    $\pm$ 0.41     & (4.82$\pm$ 0.31)$\times10^{-7}$  & 92.38    $\pm$ 28.24  & 15   & 150   & (9.06$\pm$0.53)$\times10^{-7}$& (1.09$\pm$ 0.06)$\times10^{52}$ & 1.88  & Swift/BAT  &    \\\hline
		160303A & 5     & 1.01     & -0.51   & 0         & 0.94    $\pm$ 0.13     & (9.49$\pm$ 1.28)$\times10^{-8}$ & 299.33    $\pm$ 299.17  & 15   & 150   & (3.12$\pm$0.44)$\times10^{-7}$ & (1.67 $\pm$ 0.24)$\times10^{51}$ & 3.29  & Swift/BAT  & (1,6,7)   \\\hline
		\multirow{2}{*}{160408A$\star$} & \multirow{2}{*}{0.32}  & \multirow{2}{*}{1.9}      & -0.70   & -2.66 & 12.74   $\pm$ 1.33     & (6.98$\pm$ 0.28)$\times10^{-7}$   & 841.97    $\pm$ 170.46 & 10   & 1000  & (6.53$\pm$0.68)$\times10^{-6}$ & (1.66$\pm$0.17)$\times10^{53}$& 1.54  & Fermi      & \multirow{2}{*}{(1,2,5,6,7,8,9)}   \\
		&   &        & 0.55    & 0         & 2.01    $\pm$ 0.35     & (1.99$\pm$0.34)$\times10^{-7}$  & 107.36$\pm$ 27.92  & 15  & 150  & (2.60$\pm$0.45)$\times10^{-7}$ & (6.64$\pm$1.15)$\times10^{51}$& 1.31  & Swift/BAT  &    \\\hline
		\multirow{2}{*}{160410A$\star$} & \multirow{2}{*}{8.2 }  & \multirow{2}{*}{1.7177}   & -0.71       & 0         & (2.80$\pm$0.40)$\times10^{-5}$& (1.20$\pm$0.30)$\times10^{-5}$   & 1416      $\pm$ 356      & 20   & 10000 & (2.65$\pm$0.38)$\times10^{-8}$ & (5.27$\pm$0.75)$\times10^{53}$& 0.94 & Konus-Wind & \multirow{2}{*}{(1,3,6,7,9)}   \\
		&     &     & -0.01 & 0         & 0.34    $\pm$ 0.21     & (3.13$\pm$ 1.81)$\times10^{-8}$ & 108.23    $\pm$ 108.23  & 15   & 150   & (4.27$\pm$  2.61)$\times10^{-8}$& (8.51$\pm$5.19)$\times10^{50}$& 1.37  & Swift/BAT  &    \\\hline
		\multirow{2}{*}{160411A} & \multirow{2}{*}{0.36}  &\multirow{2}{*}{ 0.82}     & 0.31    & -4.95 & 5.52    $\pm$ 1.39     & (2.25$\pm$0.19)$\times10^{-7}$  & 161.53    $\pm$ 41.38 & 10   & 1000  & (8.79$\pm$  2.21)$\times10^{-7}$& (2.82$\pm$0.71)$\times10^{51}$ & 1.02  & Fermi      & \multirow{2}{*}{(1,2,6,7,9)}  \\
		&    &       & 1.24     & 0         & 1.37    $\pm$ 0.35     & (9.13$\pm$ 2.48)$\times10^{-8}$ & 51.93    $\pm$ 10.14  & 15   & 150   & (9.67$\pm$  2.47)$\times10^{-8}$& (3.10$\pm$0.79)$\times10^{50}$ & 1.06  & Swift/BAT  &    \\\hline
		160525B$\star$ & 0.29  & 0.64     & -1.46    & 0         & 0.89    $\pm$ 0.17     & (6.92$\pm$ 1.48)$\times10^{-8}$ & 55.63     $\pm$  36.11 $^{c}$        & 15   & 150   & (1.00$\pm$0.19)$\times10^{-7}$ & (1.75$\pm$0.34)$\times10^{50}$ & 1.78  & Swift/BAT  & (1,6,7,8)  \\\hline
		160601A & 0.12  & 1.615 $^{b}$    & -0.45   & 0         & 0.90     $\pm$ 0.19     & (8.86$\pm$ 1.87)$\times10^{-8}$ & 214.22   $\pm$ 214.49  & 15   & 150   & (2.10$\pm$0.45)$\times10^{-7}$ & (3.59$\pm$0.77)$\times10^{51}$& 2.37  & Swift/BAT  & (1,5,6,7)   \\\hline
		\multirow{2}{*}{160624A$\star$} & \multirow{2}{*}{0.2}   & \multirow{2}{*}{0.4842}   & -0.63  & -3.64 & 7.43    $\pm$ 1.02     & (3.92$\pm$ 0.08)$\times10^{-7}$   & 1168.39   $\pm$ 546.52 & 10   & 1000  & (5.54$\pm$0.76)$\times10^{-6}$ & (4.93$\pm$0.68)$\times10^{51}$& 1.89  & Fermi      & \multirow{2}{*}{(1,2,4,6,7,8,9)}   \\
		&     &     & -0.54   & 0         & 0.51    $\pm$ 0.13     & (5.62$\pm$ 0.00)$\times10^{-8}$ & 52.47    $^{c}$         & 15   & 150   &( 4.00$\pm$  1.05)$\times10^{-8}$& (3.63$\pm$ 0.94)$\times10^{49}$ & 1.26  & Swift/BAT  &    \\\hline
		\multirow{2}{*}{160821B$\star$} & \multirow{2}{*}{0.48}  & \multirow{2}{*}{0.1619}   & -0.76  & -2.01 & 9.16    $\pm$ 1.19     & (1.95$\pm$0.20)$\times10^{-7}$   & 38.17     $\pm$ 27.48 & 10   & 1000  & (1.26$\pm$0.16)$\times10^{-6}$ & (9.10$\pm$ 1.18)$\times10^{49}$& 3.41  & Fermi      & \multirow{2}{*}{(1,2,6,7,8,9)}   \\
		&    &     & -0.16   & 0         & 1.68    $\pm$ 0.20      & (1.07$\pm$0.15)$\times10^{-7}$  & 50.98$\pm$7.44  & 15   & 150   & (1.27$\pm$0.15)$\times10^{-7}$ & (9.12$\pm$1.06)$\times10^{48}$ & 1.18  & Swift/BAT  &    \\  \hline
		160927A & 0.48  & 0.406 $^{b}$    & -1.28     & 0         & 0.85    $\pm$ 0.43     & (6.02$\pm$ 1.26)$\times10^{-8}$ & 53.48     $\pm$  34.49 $^{c}$       & 15   & 150   & (8.52$\pm$  4.27)$\times10^{-8}$& (4.99$\pm$ 2.50)$\times10^{49}$& 1.60   & Swift/BAT  & (1,6,7)   \\\hline
		\multirow{2}{*}{161001A} & \multirow{2}{*}{2.6}   & \multirow{2}{*}{0.67}     & -0.94   & -4.36 & 17.61   $\pm$ 1.25     & (1.76$\pm$0.02)$\times10^{-6}$ & 372.61    $\pm$ 59.60 & 10   & 1000  & (3.62$\pm$0.26)$\times10^{-6}$   & (7.08$\pm$0.50)$\times10^{51}$ & 1.11  & Fermi      & \multirow{2}{*}{(1,2,6,7,9)}   \\
		&     &       & -0.57   & 0         & 3.41    $\pm$ 0.24     & (3.46$\pm$0.26)$\times10^{-7}$  & 375.70    $\pm$ 375.70  & 15   & 150   & (1.42$\pm$0.10)$\times10^{-6}$ & (2.78$\pm$0.20)$\times10^{51}$ & 4.11  & Swift/BAT  &    \\\hline
		161104A & 0.1   & 0.793    & -1.29    & 0         & 0.59    $\pm$ 0.13     & (4.87$\pm$ 1.07)$\times10^{-8}$& 50.42     $\pm$  32.99 $^{c}$        & 15   & 150   & (5.88$\pm$  1.30)$\times10^{-8}$& (1.74$\pm$0.39)$\times10^{50}$ & 1.62  & Swift/BAT  & (1,6,7)   \\\hline
		\multirow{2}{*}{170127B$\star$} & \multirow{2}{*}{0.51}  & \multirow{2}{*}{2.2}      & -0.29  & -2.15 & 8.44    $\pm$ 1.24     & (3.06$\pm$0.13)$\times10^{-7}$   & 481.68$\pm$ 131.52 & 10   & 1000  & (4.28$\pm$0.63)$\times10^{-6}$ & (1.56$\pm$0.23)$\times10^{53}$ & 1.56  & Fermi      & \multirow{2}{*}{(1,2,6,7,8,9)}   \\
		&    &        & 0.96    & 0         & 1.07    $\pm$ 0.20      & (1.05$\pm$0.22)$\times10^{-7}$   & 95.79     $\pm$ 23.61  & 15   & 150   & (1.25$\pm$ 0.23)$\times10^{-7}$ & (4.58$\pm$0.85)$\times10^{51}$ & 1.20   & Swift/BAT  &    \\\hline
		\multirow{2}{*}{170428A$\star$} & \multirow{2}{*}{0.2}   & \multirow{2}{*}{0.453}    & -0.47       & -2.46     & (5.00$\pm$ 1.52)$\times10^{-5}$ & (4.20$\pm$0.90)$\times10^{-6}$   & 982       $\pm$ 355      & 20   & 10000 & (6.25$\pm$ 1.90)$\times10^{-5}$ & (4.75$\pm$ 1.44)$\times10^{52}$ & 1.25  & Konus-Wind & \multirow{2}{*}{(1,3,5,6,7,9)}   \\
		&     &      & -0.58   & 0         & 2.80     $\pm$ 0.26     & (2.83$\pm$0.28)$\times10^{-7}$  & 371.96    $\pm$ 371.96  & 15   & 150   & (1.14$\pm$0.10)$\times10^{-6}$& (8.66$\pm$0.79)$\times10^{50}$ & 4.03  & Swift/BAT  &    \\\hline
		170728A$\star$ & 1.25  & 1.493    & 1.15    & 0         & 1.01    $\pm$ 0.18     & 0  & 40.99     $\pm$ 5.97  & 15   & 150   & (6.26$\pm$  1.09)$\times10^{-8}$ & (8.83 $\pm$ 1.54)$\times10^{50}$ & 1.10   & Swift/BAT  & (1,7,9)   \\\hline
		\multirow{3}{*}{170728B} & \multirow{3}{*}{47.7}  & \multirow{3}{*}{1.27}     & -0.96  & -2.43  & 23.06   $\pm$ 1.43     & (4.02$\pm$0.03)$\times10^{-6}$   & 136.48    $\pm$ 28.70 & 10   & 1000  & (3.19$\pm$0.20)$\times10^{-6}$ & (3.01$\pm$0.19)$\times10^{52}$& 1.45  & Fermi      & \multirow{3}{*}{(1,2,3,6,7,9)}   \\
		&    &     & -0.22       & 0         & (4.38$\pm$0.91)$\times10^{-6}$& (3.96$\pm$0.75)$\times10^{-6}$   & 166      $\pm$ 22       & 20   & 10000 & (4.54$\pm$0.94)$\times10^{-6}$ & (4.29$\pm$ 0.89)$\times10^{52}$ & 1.04  & Konus-Wind &    \\
		&   &      & 0.48    & 0         & 7.37    $\pm$ 1.31     & (6.70$\pm$ 1.10)$\times10^{-7}$ & 91.71     $\pm$ 19.85   & 15   & 150   & (8.09$\pm$ 1.44)$\times10^{-7}$ & (7.65$\pm$ 1.36)$\times10^{51}$& 1.21  & Swift/BAT  &    \\    \hline
		170817A$\ast$ & 2     & 0.009787 & 0.15   & -8.94 & 3.73    $\pm$ 0.93     & (2.79$\pm$0.17)$\times10^{-7}$   & 214.70     $\pm$ 56.59 & 10   & 1000  & (7.33$\pm$ 1.82)$\times10^{-7}$ & (1.57 $\pm$0.39)$\times10^{47}$& 1.00     & Fermi      & (1,2,4,5,7,9)   \\\hline
		\multirow{2}{*}{180418A$\star$} & \multirow{2}{*}{2.29}  & \multirow{2}{*}{1.56}     & -1.38   & -9.41 & 7.56    $\pm$ 1.14     & (5.90$\pm$0.09)$\times10^{-7}$   & 1051.09   $\pm$ 951.48 & 10   & 1000  &( 1.95$\pm$0.29)$\times10^{-6}$  & (3.06$\pm$0.46)$\times10^{52}$ & 1.57  & Fermi      & \multirow{2}{*}{(1,2,6,7,9)}   \\
		&    &      & -1.40   & 0         & 3.02    $\pm$ 0.19     & (2.39$\pm$0.17)$\times10^{-7}$  & 78.68      $\pm$ 37.51 $^{c}$          & 15   & 150   & (3.62$\pm$0.23)$\times10^{-7}$ & (5.68 $\pm$ 0.35) $\times10^{51}$& 1.79  & Swift/BAT  &    \\\hline
		\multirow{3}{*}{180618A$\star$} & \multirow{3}{*}{47.4}  & \multirow{3}{*}{0.52}     & 1.08        & 0         & 14.60    $\pm$ 1.90      & (2.12$\pm$0.12)$\times10^{-6}$   & 2800      $\pm$ 800      & 10   & 1000  & (1.55 $\pm$ 0.20)$\times10^{-4}$ & (1.63$\pm$0.21)$\times10^{53}$ & 10.80  & Fermi      & \multirow{3}{*}{(1,2,3,6,7,9)}   \\
		&    &      & -0.36       & 0         &( 5.60$\pm$ 1.83)$\times10^{-5}$ & (6.24$\pm$ 1.20)$\times10^{-6}$& 2461       $\pm$ 565      & 20   & 10000 & (5.40$\pm$ 1.77)$\times10^{-5}$ & (5.71 $\pm$ 1.87)$\times10^{52}$ & 0.96 & Konus-Wind &    \\
		&    &      & -0.97  & 0         & 2.15    $\pm$ 0.22     &( 2.04$\pm$0.22)$\times10^{-7}$   & 75.27$\pm$ 40.39 $^{c}$          & 15   & 150   & (2.17$\pm$0.22)$\times10^{-7}$ & (2.30 $\pm$ 0.23)$\times10^{50}$ & 1.44  & Swift/BAT  &    \\\hline
		\multirow{2}{*}{180727A} & \multirow{2}{*}{1.1}   & \multirow{2}{*}{1.95}     & 0.14        & 0         & 7.20     $\pm$ 1.20      & (3.32$\pm$0.18)$\times10^{-7}$  & 69  $\pm$ 4        & 10   & 1000  & (5.37$\pm$0.89)$\times10^{-7}$ & (1.46$\pm$0.24)$\times10^{52}$ & 1.03  & Fermi      & \multirow{2}{*}{(1,2,5,6,7,9)}   \\
		&     &       & -0.45    & 0         & 3.50     $\pm$ 0.24     & (2.53$\pm$0.21)$\times10^{-7}$  & 69.68     $\pm$ 9.47  & 15   & 150   & (3.15$\pm$0.21)$\times10^{-7}$& (8.57$\pm$0.58)$\times10^{51}$& 1.25  & Swift/BAT  &    \\\hline
		\multirow{2}{*}{180805B$\star$} & \multirow{2}{*}{122.5} & \multirow{2}{*}{0.6612}   & -0.50        & 0         & 8.30     $\pm$ 1.10      & (5.90$\pm$0.70)$\times10^{-7}$   & 346       $\pm$ 75       & 10   & 1000  & (2.00$\pm$0.27)$\times10^{-6}$ & (3.79$\pm$0.50)$\times10^{51}$ & 1.04  & Fermi      & \multirow{2}{*}{(1,2,6,7,9)}  \\
		&   &    & 0.76    & 0         & 1.61    $\pm$ 0.23     & (1.66$\pm$0.24)$\times10^{-7}$   & 111.25    $\pm$ 26.25  & 15   & 150   & (2.20$\pm$0.32)$\times10^{-7}$ & (4.17$\pm$0.61)$\times10^{50}$ & 1.33  & Swift/BAT  &    \\\hline
		181123B$\star$ & 0.26  & 1.754    & -0.62    & 0         & 1.32    $\pm$ 0.25     & (1.43$\pm$ 0.00)$\times10^{-7}$  & 68.17  $^{c}$         & 15   & 150   & (1.20$\pm$0.23)$\times10^{-7}$ & (2.52$\pm$0.48)$\times10^{51}$& 1.29  & Swift/BAT  &(1,5,6,7,8)   \\\hline
		\multirow{3}{*}{191031D} & \multirow{3}{*}{0.29}  & \multirow{3}{*}{1.93}     & -0.42       & 0         & 48.70    $\pm$ 1.59     &( 4.36$\pm$0.09)$\times10^{-6}$  & 856        $\pm$ 53       & 10   & 1000  & (2.96$\pm$0.10)$\times10^{-5}$& (7.86$\pm$0.26)$\times10^{53}$ & 1.47  & Fermi      & \multirow{3}{*}{(1,2,3,6,7,9)}   \\
		&    &      & -0.27       & 0         & (1.95$\pm$0.32)$\times10^{-5}$ & (2.72$\pm$0.27)$\times10^{-6}$  & 370  $\pm$ 46       & 20   & 10000 & (1.97$\pm$0.32)$\times10^{-5}$ & (5.22$\pm$ 0.86)$\times10^{53}$ & 1.01  & Konus-Wind &    \\
		&    &      & -0.86   & 0         & 4.24    $\pm$ 0.40      & (4.19$\pm$ 0.00)$\times10^{-7}$  & 92.09   $^{c}$      & 15   & 150   & (4.66$\pm$0.44)$\times10^{-7}$ & (1.24$\pm$ 0.12)$\times10^{52}$& 1.46  & Swift/BAT  &    \\\hline
		\multirow{3}{*}{200219A$\star$} & \multirow{3}{*}{288 }  & \multirow{3}{*}{0.48}     & -0.60        & 0         & 16.10    $\pm$ 1.10      & (2.80$\pm$0.10)$\times10^{-6}$   & 1400       $\pm$ 200      & 10   & 1000  & (1.46$\pm$0.10)$\times10^{-5}$ & (1.28$\pm$0.09)$\times10^{52}$& 2.15  & Fermi      & \multirow{3}{*}{(1,2,3,6,7,9)}   \\
		&     &       & -0.06       & 0         & (2.34$\pm$0.73)$\times10^{-5}$ & (4.12$\pm$0.61)$\times10^{-6}$   & 952        $\pm$ 173      & 20   & 10000 & (2.34$\pm$0.73)$\times10^{-5}$ &(2.04 $\pm$0.64)$\times10^{52}$ & 1.00     & Konus-Wind &    \\
		&     &       & 1.00           & 0         & 0.06    $\pm$ 0.11     & (10.1$\pm$ 7.98)$\times10^{-9}$  & 80         $\pm$ 80       & 15   & 150   & (5.88$\pm$0.11)$\times10^{-9}$ & (5.13 $\pm$ 9.13)$\times10^{48}$ & 1.11  & Swift/BAT  &    \\\hline
		\multirow{2}{*}{200411A} & \multirow{2}{*}{0.22}  & \multirow{2}{*}{0.7}      & -0.64       & 0         & 11.90    $\pm$ 1.50      &( 4.70$\pm$0.60)$\times10^{-7}$   & 420        $\pm$ 120     & 10   & 1000  & (3.17$\pm$ 0.40)$\times10^{-6}$ & (6.88$\pm$0.87)$\times10^{51}$ & 1.09  & Fermi      & \multirow{2}{*}{(1,2,6,7,9)}   \\
		&    &       & 0.44    & 0         & 1.07    $\pm$ 0.17     & (1.02$\pm$0.19)$\times10^{-7}$  & 178.22     $\pm$ 178.24  & 15   & 150   & (2.60$\pm$0.41)$\times10^{-7}$ & (5.65$\pm$0.90)$\times10^{50}$& 2.10   & Swift/BAT  &    \\\hline
		200522A & 0.62  & 0.5536   & 0.60    & 0         & 1.46    $\pm$ 0.17     & (1.11$\pm$0.16)$\times10^{-7}$   & 84.09      $\pm$ 23.18  & 15   & 150   & (1.48$\pm$0.18)$\times10^{-7}$ & (1.82 $\pm$0.22)$\times10^{50}$ & 1.16  & Swift/BAT  & (1,6,7,9)   \\\hline
		\multirow{2}{*}{200826A$\ast$} & \multirow{2}{*}{1.14}  & \multirow{2}{*}{0.7481}   & -0.41       & -2.40      & 39.06   $\pm$ 0.42     & (4.80$\pm$0.10)$\times10^{-6}$  & 89.80       $\pm$ 3.70     & 10   & 1000  & (5.15$\pm$0.06)$\times10^{-6}$ & (1.32$\pm$0.01)$\times10^{52}$ & 1.68  & Fermi      & \multirow{2}{*}{(2,3,9)}   \\
		&    &     & 1.27        & -2.32     & (9.04$\pm$ 2.49)$\times10^{-6}$ & (3.30$\pm$0.42)$\times10^{-6}$  & 67         $\pm$ 17       & 20   & 10000 & (2.17$\pm$0.60)$\times10^{-5}$ & (5.55 $\pm$ 1.53)$\times10^{52}$ & 2.40   & Konus-Wind &    \\\hline
		200907B & 0.83  & 0.56     & -0.62     & 0         & 1.51    $\pm$ 0.20      & (1.63$\pm$0.21)$\times10^{-7}$  & 70.65       $\pm$  39.89 $^{c}$         & 15   & 150   & (1.40$\pm$0.19)$\times10^{-7}$ & (1.77$\pm$0.23)$\times10^{50}$ & 1.29  & Swift/BAT  & (1,6,7)   \\\hline
		\multirow{2}{*}{201006A} & \multirow{2}{*}{0.49}  & \multirow{2}{*}{0.299 $^{b}$}    & -0.97       & 0         & 6.89    $\pm$ 0.97     & (3.33$\pm$0.38)$\times10^{-7}$   & 130         $\pm$ 26       & 10   & 1000  & (6.95$\pm$0.98)$\times10^{-7}$ & (1.99$\pm$0.28)$\times10^{50}$& 1.07  & Fermi      & \multirow{2}{*}{(1,2,6,7)}   \\
		&    &     & 1.37     & 0         & 2.02    $\pm$ 0.35     & (1.37$\pm$0.27)$\times10^{-7}$  & 52.96      $\pm$ 7.74   & 15   & 150   & (1.44$\pm$0.25)$\times10^{-7}$ & (4.13$\pm$0.71)$\times10^{49}$ & 1.05  & Swift/BAT  &    \\\hline
		\multirow{3}{*}{201221D} & \multirow{3}{*}{0.16}  & \multirow{3}{*}{1.055}    & 0.01        & -3.30      & 41     $\pm$ 2        & (1.08$\pm$0.05)$\times10^{-6}$   & 98         $\pm$ 8       & 10   & 1000  & (4.45$\pm$0.22)$\times10^{-6}$& (2.66$\pm$0.13)$\times10^{52}$ & 1.17  & Fermi      & \multirow{3}{*}{(1,2,3,6,7,9)}   \\
		&    &     & -0.95       & 0         & (6.60$\pm$ 1.30)$\times10^{-6}$ & (5.30$\pm$ 1.10)$\times10^{-7}$  & 148        $\pm$ 37      & 20   & 10000 & (7.45$\pm$1.47)$\times10^{-6}$ & (4.45 $\pm$ 0.88)$\times10^{52}$ & 1.13  & Konus-Wind &    \\
		&    &      & -0.99   & 0         & 5.47    $\pm$ 0.55     & (4.04$\pm$0.47)$\times10^{-7}$   & 93.98      $\pm$ 28.64  & 15   & 150   &( 6.18$\pm$0.62)$\times10^{-7}$ & (3.69$\pm$0.37)$\times10^{51}$ & 1.53  & Swift/BAT  &    \\\hline
		\multirow{3}{*}{210323A} & \multirow{3}{*}{1.12}  & \multirow{3}{*}{0.733}    & -0.97       & -3.02     & 21.30    $\pm$ 1.50      & (1.20$\pm$0.10)$\times10^{-6}$   & 2100       $\pm$ 500      & 10   & 1000  & (1.69$\pm$0.12)$\times10^{-5}$ & (4.12$\pm$0.29)$\times10^{52}$ & 2.55  & Fermi      & \multirow{3}{*}{(1,2,3,6,7,9)}   \\
		&    &      & -1.03       & 0         & (1.23$\pm$0.20)$\times10^{-5}$ & (1.57$\pm$0.25)$\times10^{-6}$   & 632        $\pm$ 167     & 20   & 10000 & (1.27$\pm$0.21)$\times10^{-5}$ & (3.10 $\pm$ 0.50)$\times10^{52}$ & 1.03  & Konus-Wind &    \\
		&    &      & -1.39    & 0         & 1.42    $\pm$ 0.40      & (2.36$\pm$0.22)$\times10^{-7}$ & 78.45      $\pm$ 40.23 $^{c}$         & 15   & 150   & (1.67$\pm$0.47)$\times10^{-7}$ & (4.08 $\pm$ 1.15)$\times10^{50}$ & 1.75  & Swift/BAT  &    \\\hline
		210726A & 0.39  & 0.37     & -0.84   & 0         & 0.63    $\pm$ 0.18     & (4.13 $\pm$ 1.56)$\times10^{-8}$  & 57.75     $\pm$ 57.74  & 15   & 150   & (5.56$\pm$  1.55)$\times10^{-8}$ & (2.62$\pm$0.73)$\times10^{49}$ & 1.35  & Swift/BAT  & (1,6,7,9)   \\\hline
		210919A & 0.16  & 0.2415   & -1.14    & 0         & 6.42    $\pm$ 1.06     & (5.36$\pm$ 1.95)$\times10^{-8}$  & 107.59    $\pm$ 49.58  & 15   & 150   & (7.92$\pm$  1.31)$\times10^{-7}$ & (1.39$\pm$0.23)$\times10^{50}$ & 1.67  & Swift/BAT  & (1,6,7)   \\\hline
		211023B & 1.3   & 0.862    & -1.66    & 0         & 2.21    $\pm$ 0.32     &( 1.37$\pm$0.25)$\times10^{-7}$ & 50.93      $\pm$ 50.84  & 15   & 150   &   (2.85$\pm$0.41)$\times10^{-7}$ & (1.03 $\pm$0.15)$\times10^{51}$ & 2.09  & Swift/BAT  & (1,6,7)   \\\hline
		211106A & 1.75  & 0.097    & -0.02       & 0         & (3.37$\pm$ 1.59)$\times10^{-6}$& (6.09$\pm$ 1.22)$\times10^{-7}$ & 196        $\pm$ 45       & 20   & 10000 & (3.43$\pm$ 1.62)$\times10^{-6}$ & (8.17$\pm$ 3.85)$\times10^{49}$& 1.02  & Konus-Wind & (3,6,7,9)   \\\hline
		\multirow{2}{*}{211211A$\ast$} & \multirow{2}{*}{51.37} & \multirow{2}{*}{0.0763}   & -1.3        & -2.4      & 324.90   $\pm$ 1.50      & (5.40$\pm$0.01)$\times10^{-4}$   & 646.80      $\pm$ 7.80      & 10   & 1000  & (8.65$\pm$0.04)$\times10^{-5}$ & (1.24$\pm$0.06)$\times10^{50}$& 1.60   & Fermi      & \multirow{2}{*}{(1,2,6,7,9)}   \\
		&   &     & -0.73   & 0         & 153.09  $\pm$ 3.27     & (1.43$\pm$0.03)$\times10^{-5}$  & 260.07     $\pm$ 64.12 & 15   & 150   & (3.87$\pm$0.08)$\times10^{-5}$  & (5.54$\pm$0.12)$\times10^{50}$ & 2.71  & Swift/BAT  &    \\\hline
		\multirow{2}{*}{211227A$\ast$} & \multirow{2}{*}{83.79} &\multirow{2}{*}{ 0.228}    & -1.34       & -2.26     & (2.00$\pm$0.40)$\times10^{-6}$& (2.60$\pm$0.21)$\times10^{-5}$   & 192       $\pm$ 42       & 15   & 1500  & (3.22$\pm$0.64)$\times10^{-6}$ & (4.97$\pm$0.99)$\times10^{50}$ & 1.61  & Konus-Wind & \multirow{2}{*}{(1,3,9)}   \\
		&   &      & -1.26   & 0         & 5.05    $\pm$ 0.43     & (3.37$\pm$0.36)$\times10^{-7}$   & 69.39      $\pm$ 18.67  & 15   & 150   & (5.38$\pm$0.46)$\times10^{-7}$ & (8.31 $\pm$0.71)$\times10^{49}$& 1.60   & Swift/BAT  &    \\\hline
		\multirow{2}{*}{220617}  & \multirow{2}{*}{0.704} & \multirow{2}{*}{0.807}    & -0.70        & 0         & 10      $\pm$ 1        & (2.30$\pm$0.10)$\times10^{-6}$   & 2300       $\pm$ 300      & 10   & 1000  & (1.32$\pm$0.13)$\times10^{-5}$ & (4.06$\pm$0.41)$\times10^{52}$ & 3.09  & Fermi      & \multirow{2}{*}{(2,3)}   \\
		&   &     & 0.02        & 0         & (3.31$\pm$ 1.27)$\times10^{-5}$ & (6.53$\pm$ 1.15)$\times10^{-6}$  & 1384       $\pm$ 341      & 20   & 10000 & (3.30$\pm$ 1.27)$\times10^{-5}$& (1.02 $\pm$ 0.39)$\times10^{53}$ & 0.99 & Konus-Wind &    \\\hline
		221025  & 0.448 & 0.39     & -0.40        & 0         & 13.30    $\pm$ 0.90      & (1.70$\pm$0.10)$\times10^{-6}$   & 1680       $\pm$ 345      & 10   & 1000  & (1.85$\pm$0.13)$\times10^{-5}$ & (9.86$\pm$0.67)$\times10^{51}$ & 2.69  & Fermi      & (2)    \\
		\enddata
	\end{deluxetable*}
	\footnotesize{Note:\\ $^a$ GRB 050709 is only detected  by HETE-II.\\ b. Using the E$_{p}$-Luminosity relation, we try to estimate the redshifts of these GRBs without known redshifts \citep{2004ApJ...609..935Y,2018PASP..130e4202Z}.\\ c. We have estimated their peak energies by means of the empirical relation $ log(E_{p}) \sim 3.47+0.28 \times log(10 \times S) $ proposed by \cite{2020ApJ...902...40Z}.\\ d. The unit of the flux(P) detected by Konus-Wind is $erg \; cm^{-2}\; s^{-1}$.\\ e. The K-corrected flux: $ P_{bolo}=P \times K$.\\ f. The duration time(T$_{90}$) and fluence(S) are taken from \url{https://swift.gsfc.nasa.gov/archive/grb_table/}.\\ g. These GRBs with EE are marked with $\star$; $\ast$ represents the special type of GRBs. GRB 060614, GRB 100816A, GRB 211211A, GRB 211227A are merger-driven LGRBs; GRB 200826A is collapse-driven SGRB; GRB 170817A is the first binary neutron star merger SGRB and an off-axis burst. \\ h. Ref:(1) \url{https://swift.gsfc.nasa.gov/results/batgrbcat/}; (2) \url{https://heasarc.gsfc.nasa.gov/db-perl/W3Browse/w3table.pl?tablehead=name%3Dfermigbrst&Action=More+Options};(3) \url{https://gcn.nasa.gov/circulars};(4) \cite{2018PASP..130e4202Z};(5) \cite{2022A&A...657A.124L};(6) \cite{2022ApJ...940...56F};(7) \url{https://bright.ciera.northwestern.edu/};(8) \cite{2023ApJ...951....4G};(9) \cite{2023ApJ...950...30Z}.\\i. GRB 060614, GRB 061201, GRB 130603B, GRB 140903A and GRB 200522A are associated with kilonovae.}
\end{longrotatetable}

\begin{table*}[htbp]
	\centering
	\caption{Instrumental parameters of diffenert satellites}
	\begin{threeparttable}
	\begin{tabular}{c|c|c|c|c}
		\hline
		\hline
		Detectors  & Field of View & Operation Time  & Sensitivity&Ref \\
			& $\mathrm{\Omega (sr)}$ &  $T$(yr)  &$\mathrm{(erg\;cm^{-2}\;s^{-1}}$)&  \\
		\hline
		Swift/BAT      & 1.40             & 18             & $1.00\times10^{-08} $  & a \\
		Fermi/GBM      & 9.50                & 15             & $8.44\times10^{-08} $  & b\\
		Konus-Wind & 4$\pi$       & 18             & $1.00\times10^{-06} $    & c\\
		\hline
		\hline
	\end{tabular}%
	\label{tab:Detectors}%
	\footnotesize{Note: a. \url{https://swift.gsfc.nasa.gov/about_swift/bat_desc.html}; b. The photon flux limit of $7.1\ \rm ph\ \rm cm^{-2}\rm\ s^{-1}$ has been taken from the Fermi website (\url{https://f64.nsstc.nasa.gov/gbm/instrument/}{Fermi GBM}) for unit transformation. c. \cite{2021ApJ...908...83T}}
    \end{threeparttable}
\end{table*}%

\section{Methods} \label{sec:three}

\subsection{Time delay models} \label{subsec:HR distribution}
With the accumulation of the SFR observation data, people proposed some empirical or theoretical models to depict how the SFR evolves with the redshift \citep{1959ApJ...129..243S,2001ApJ...548..522P,2004ApJ...615..209H,2006ApJ...647..787T,2008ApJ...683L...5Y,2014ARA&A..52..415M}. Here, the analytical model given by \cite{2008ApJ...683L...5Y} is applied in our study, which simply reads
\begin{equation}
	\dot{\rho}_{\ast }\left ( z \right )    \propto \left [ \left ( 1+z \right )^{a\eta }  +\left ( \frac{1+z}{B}  \right )^{b\eta }  +\left ( \frac{1+z}{C}  \right )^{c\eta }  \right ] ^{\frac{1}{\eta } } ,
	\label{SFR-nodelay}
\end{equation}
where $\eta=-10, a=3.4, b=-0.3, c=-3.5, B=5000$ and $C=9$ obtained by the contraint of bright\textit{ Swift} LGRBs on the cosmic SFR have been adopted for our calculations. Moreover, we investigate the undelayed SFR model in Equation (15) of \cite{2014ARA&A..52..415M} as well as its diversely corresponding delay SFR models for comparation. 

Unlike the core-collapse process of massive stars, the binary star merger undergo as relatively prolonged inspiral process during the coalescence of binary stars. Therefore, an additional time delay should be considered for SGRBs with binary merger progenitors. Follow \cite{2021MNRAS.501..157Z}, we focus on  three types of merger delay time-scale ($\mathrm{\tau}$) distributions. The first one is the Gaussian merger delay time-scale model. The $\tau$ obeys probability intensity distribution as :
\begin{equation}\label{eq:G}
	P\left ( \tau  \right ) = \frac{1}{\sqrt{2\pi } \sigma } \exp\left ( -\frac{\left (  \tau -\tau_{0}   \right )^{2}  }{2\sigma ^{2} }  \right ),
\end{equation}
where $\tau_{0} = 2 \; \mathrm{Gyr}$, and $\sigma = 0.3\; \mathrm{Gyr}$ \citep{2011ApJ...727..109V}.

The second one is the log-normal merger delay time-scale model written as
\begin{equation}\label{eq:log}
    P\left ( \tau  \right ) = \frac{1}{\sqrt{2\pi } \sigma } \exp\left ( -\frac{\left (  \ln_{}{\tau}-\ln \tau_{0} \right )^{2}  }{2\sigma ^{2} }  \right ) ,
\end{equation}
where $\tau_{0} = 2.9 \; \mathrm{Gyr}$, and $\sigma = 0.2 \; \mathrm{Gyr}$ \citep{2015MNRAS.448.3026W}.

The third one is the power-law merger delay time-scale model like
\begin{equation}\label{eq:pl}
	P\left ( \tau  \right ) = \tau ^{-\alpha _{\tau } } ,
\end{equation}
where the power-law index approximates $\alpha _{\tau }=0.81$ \citep{2015MNRAS.448.3026W}. Notably, the power-law forms was found to satisfy most of SGRBs \citep[e.g.][]{2018MNRAS.477.4275P,2022MNRAS.515.4890O}. However, \cite{2021ApJ...917...24Z} pointed out that the first two $\tau$ distributions are more preferred in view of the observability of SGRBs. In other words, the real $\tau$ distribution is still ambiguous, which motivates us to utilize all the three $\tau$ distributions to investigate their corresponding SGRB rates and check which one is more coincident with the observations. To do so, we need derive the redshift distribution of SGRBs in each delay time model.

\subsection{Distributions of redshift and luminosity }
The dimensionless redshift distribution factor f(z) is associated with the GRB event rate by
\begin{equation}
	 \dot{\rho}( z ) = \dot{\rho _{0}} \cdot f(z) ,
	 \label{GRBrate-theory}
\end{equation} 
 in which $\rho _{0}$ represents the local GRB event rate at $z=0$. Considering the above three time delay models, the event rate of SGRBs should be modified by the time delay and written as
\begin{equation}
	\dot{\rho}  _{SGRB}(z)\propto  f_{SGRB}(z) =\int_{z_{min}}^{\infty} \dot{\rho}_{\ast} ( z^{'})P(\tau[z,z^{'}]) \frac{\mathrm{d}\tau}{\mathrm{d} z^{'}}\mathrm{d}z^{'}  ,
	\label{GRBrate-delay}
\end{equation}
where $\dot{\rho}_{\ast}$ denotes the SFR as shown in Eq. (\ref{SFR-nodelay}). $P(\tau)$ is the delay time-scale probability distribution function. $\tau=t (z^{'})-t(z)$ is the delay time between the formation and the merger of binary star systems. $t(z)$ is the cosmological lookback time defined by $t(z)=H_{0}^{-1}\int_{0}^{z}  \left[( 1+z ) \sqrt{\Omega _{m}( 1+z) ^{3} +\Omega _{\Lambda } }\; \right]^{-1}\;dz$, and $ t(z_{min})-t(z) = \tau_{min} = 10\; \mathrm{Myr}$ \citep{2018MNRAS.477.4275P}.

Using the three merging delay models, \cite{2015ApJ...812...33S} obtained the empirical formulas of the dimensionless redshift distribution $f(z)$. \cite{2021ApJ...917...24Z} also used the convolution but updated formule of $f(z)$ to fit the redshift distributions of the merged binary systems. However, how the SGRB event rates of different time-delay models evolve with the redshift or which one is more close to the star formation history is still unknown yet. The prompt $\gamma$-ray luminosity is calculated by $L=4\pi D_{L}^{2} F_{p} k$ , where $D_{L}$ is the luminosity distanceat a redshift of $z$. The k-correction parameter is calculated by
\begin{equation}
	k=\int_{1/(1+z)}^{10^{4}/(1+z)} EN_{E}(E)dE \times\left(\int_{e_{1}}^{e_{2}} EN_{E}(E)dE \right)^{-1}
\end{equation}
where $N_{E}(E)$ in units of $ph/cm^2/s/keV$ is the rest-frame photon spectrum of a given GRB, which is generally described by the Band function \citep{1993ApJ...413..281B}, $e_1$ and $e_2$ are the minimum and maximum value of a detector energy band.
The observed $L$ distributions can be fitted by a smoothly broken power-law (BPL) form
\begin{equation}\label{eq:BPL}
	\Phi \left ( L \right ) \propto \left [ \left ( \frac{L}{L_{b} }  \right ) ^{\omega \alpha _{1}  } + \left ( \frac{L}{L_{b} }  \right ) ^{\omega \alpha _{2}  }\right ] ^{-\frac{1}{\omega } } ,
\end{equation}
where $\alpha _{1}$ and $\alpha _{2}$ are the PL indices before and after the break luminosity ($L_{b}$), and $\omega$ is a smooth parameter characterizing the sharpness of the break.
Note that the differential luminosity satisfies $\displaystyle\int_{0}^{+\infty} \Phi(L)dL=N$ that will be used to reckon the GRB rate in the following.
\subsection{The observed event rate of SGRBs}
The differential number of SGRBs detected by a telescope with the fov of $\Omega$ among the lifetime of $T$ can be expressed by
\begin{equation}
    \begin{aligned}
	\frac{d N}{d z} = \frac{\Omega T}{4\pi } \frac{f\left ( z \right ) }{
		1+z} \frac{d V}{d z}\int_{L} \dot{\rho} _{0L}dL ,
    \end{aligned}
\label{dNdz}
\end{equation}
in which the specific SGRB event rate density at a given $L$ can be written as $\rho_{0L}=\dot{\rho _{0}}\Phi(L)$, here $\rho_{0}$ stands for the local SGRB rate with luminosity larger than $L_{min}$. $\frac{dV}{dz}$ is the differential comoving volume. Substituting $\rho_{0L}$ into Eq. (\ref{dNdz}), one can obtain the following form within the $L$ interval ranging from $L_{min}$ to $L_{max}$ as
\begin{equation}
	\begin{aligned}
		\frac{d N}{d z} =\frac{\Omega T}{4\pi } \frac{f\left ( z \right ) }{1+z} \frac{d V}{d z}\int_{L_{min} }^{L_{max} } \dot{\rho _{0}} \Phi \left ( L \right ) dL ,
	\end{aligned}
\end{equation}
According to Eq. (\ref{GRBrate-theory}), we have the redshift-dependent event rate density of $R_{SGRB}\equiv\dot{\rho}(z)=\dot{\rho_0}f(z)$.  
Thus the above formula can be converted into the observed redshift-dependent SGRB rate to be
\begin{equation}\label{R}
	R_{SGRB}=\frac{\mathrm{d} N}{\mathrm{d} z}\frac{4\pi }{\Omega T}(1+z)\left(\frac{\mathrm{d} V}{\mathrm{d} z} \right)^{-1} \left[\int_{L_{min}}^{L_{max}}  \Phi( L) \mathrm{d}L \right]^{-1}.
\end{equation}

\section{result} \label{sec:four}
\subsection{Differential distributions of redshift and Luminosity }
\begin{figure}
	\setlength{\abovecaptionskip}{-0.8cm} 
	\gridline{\fig{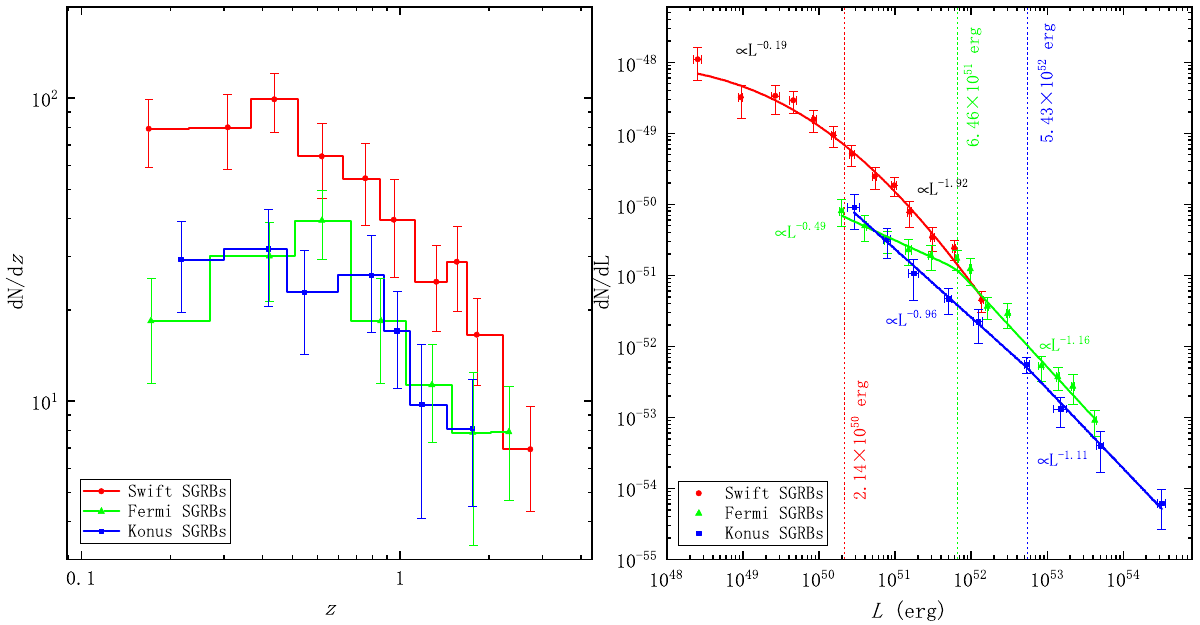}{\textwidth}{}}
	\caption{Differential redshift (left panel) and luminosity (right panel) distribution of 95 Swift (red), 41 Fermi (green) and 32 Konus-Wind (blue) SGRBs. Poisson errors have been given to each data point by the error propagation. The solid lines on the right panel  mark the best fits to observations with Eq. (\ref{eq:BPL}). \label{fig:z}}
\end{figure}

 Figure \ref{fig:z} shows the redshift and luminosity distribution of SGRBs detected by different detectors. It can be seen that the fraction of Swift SGRBs with lower redshift or smaller luminosity is significantly larger than that of Fermi or Konus-Wind SGRBs. While the redshift/luminosity distributions of Fermi and Konus-Wind SGRBs are quite analogous. This indicates that the distributions of both redshift and luminosity are obviously influenced by the energy band of detectors. We utilize the BPL model in Eq. (\ref{eq:BPL}) respectively  to fit the $L$ distributions and list our results in Table \ref{tab:L}. The $L_{b}$ value of Swift SGRBs we obtain is about $2.14\times 10^{50} \;\mathrm{erg \; s^{-1}}$ that is similar to some previous estimates \citep[e.g.][]{2018ApJ...852....1Z} which is roughly one order of magnitude less than that of LGRBs \cite[e.g.][]{2018PASP..130e4202Z,2022MNRAS.513.1078D,2023ApJ...958...37D}. 
 However, it is interestingly found that the $L_b$ value of Fermi SGRBs is comparable to those of the traditional LGRBs, but larger/smaller than that of Swift/Konus SGRBs about one order of magnitude, implying that the $L_b$ measurement is also affected by the band width of detectors and both types of bursts could share the same radiation mechanism. 
\begin{table*}[htbp]
	\centering
	\caption{The Best-fitting parameters of the differential $L$ distributions }
	\begin{tabular}{c|c|c|c|c|c}
		\hline
		\hline
		Detectors  & Model & $\alpha_{1}$  & $\alpha_{2}$ & $L_{b}$ ($\rm erg \rm s^{-1}$)& $\chi^{2}/dof$\\
		\hline
		Swift/BAT      & BPL    &  -0.19$\pm$0.08   & 1.92$\pm$0.06  & $(2.14\pm0.21)\times 10^{50}$& 1.62\\
		Fermi/GBM      & BPL    & 0.49$\pm$0.05 &1.16$\pm$0.03 &$(6.46\pm0.34)\times 10^{51}$  & 2.09\\
		Konus-Wind &BPL        & 0.96$\pm$0.04   & 1.11$\pm$0.07  &  $(5.43\pm0.54)\times 10^{52}$ &1.73\\
		\hline
		\hline 
	\end{tabular}%
	\label{tab:L}%
\end{table*}%
\subsection{Modelling the time-delayed reshift distribution functions}
\begin{figure}
	\gridline{\fig{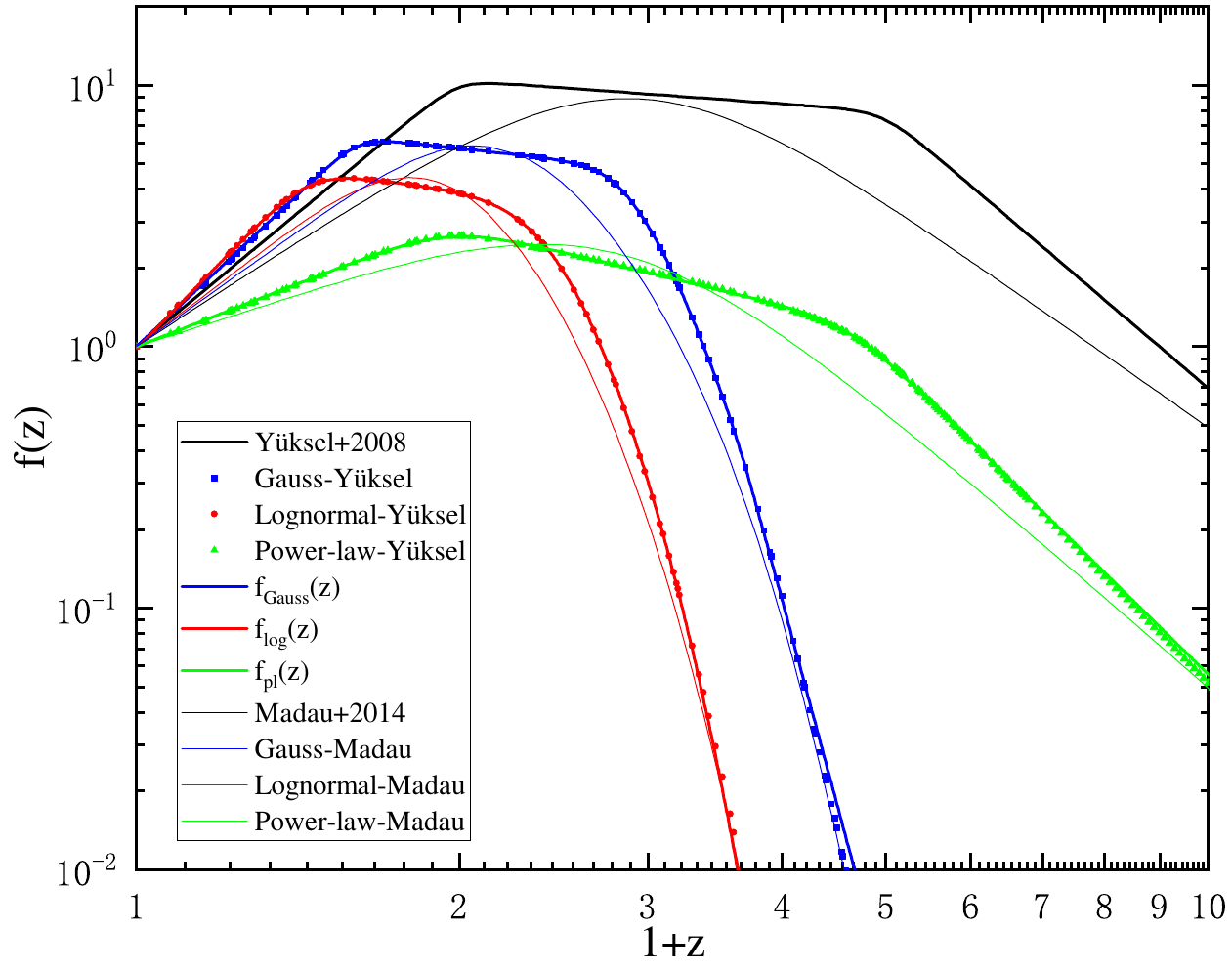}{0.75\textwidth}{}}
	\caption{Redshift distributions of differently delayed and undelayed SFRs symbolized with thick \citep{2008ApJ...683L...5Y} and thin \citep{2014ARA&A..52..415M} lines. All the distributions have been normalized to unity in the local universe ($z = 0$). The colorful thick curves represent the best fits with the merged delay models in each. The black curves are given by the undelayed SFRs. \label{fig:fz}}
\end{figure}
Once the above three time delay models are applied, the expected event rates of SGRBs in the framework of binary star mergers might vary to a large extent. Particularly, they may not coordinate with the traditional star formation history. In terms of Eqs. (\ref{GRBrate-theory}) and (\ref{GRBrate-delay}), we need to obtain the redshift distributions $f(z)$ of the three cases of time-delay models on basis of the undelayed \cite{2008ApJ...683L...5Y} model. For this, we substitute Eqs. (2)-(4) into Eq. (\ref{GRBrate-delay}) and calculate the empirical redshift distribution functions separately. Assuming the deduced $f(z)$ in each model case has the similar form as Eq. (\ref{SFR-nodelay}),  we then follow \cite{2021ApJ...917...24Z} to fit the empirical redshift distributions in Figure \ref{fig:fz} and get the redshift-dependent distribution functions as
\begin{equation}\label{fz1}
\begin{aligned}
	f_{Gauss}(z)=&\left[ (1+z)^{3.86\eta_{G} }+\left(\frac{1+z}{107.17}\right)^{-0.44\eta_{G} }+\left(\frac{1+z}{3.54}\right)^{-6.65\eta_{G} }\right. \\&\left. +\left(\frac{1+z}{3.39}\right)^{-10.51\eta_{G} }+\left(\frac{1+z}{3.46}\right)^{-15.26\eta_{G} }+\left(\frac{1+z}{1.50}\right)^{-3.17\eta_{G} }  \right]^{1/\eta_{G} },\\
\end{aligned}
\end{equation}
for the Guassian time-delay model, 
\begin{equation}\label{fz2}
	\begin{aligned}
	f_{log}(z)=&\left[(1+z)^{4.23\eta_{log} }+\left(\frac{1+z}{25.84}\right)^{-0.54\eta_{log} }+\left(\frac{1+z}{3.25}\right)^{-3.52\eta_{log} }+\left(\frac{1+z}{3.00}\right)^{-6.07\eta_{log} }\right.\\&\left.+\left(\frac{1+z}{0.99}\right)^{374514.58\eta_{log} }+\left(\frac{1+z}{2.71}\right)^{-\left ( 0.31+\frac{z^{1.05} }{0.18}  \right ) ^{\eta_{log} }} \right]^{1/\eta_{log} },\\
\end{aligned}
\end{equation}
for the log-normal time-delay model,  and
\begin{equation}\label{fz3}
	\begin{aligned}
		f_{pl}(z)=&\left[(1+z)^{1.59\eta_{pl} }+\left(\frac{1+z}{4.88}\right)^{-4.03\eta_{pl} }+\left(\frac{1+z}{5.79}\right)^{-0.99\eta_{pl} }\right]^{1/\eta_{pl} },
	\end{aligned}
\end{equation}
for the power-law time-delay model,
where $\eta_{G}=-9.39\pm0.22$, $\eta_{log}=-6.26\pm0.49$ and $\eta_{pl}=-6.39\pm0.13$. In addition, we plot the redshift distributions of the delayed and undelayed SFRs according to the \cite{2014ARA&A..52..415M} model in Figure \ref{fig:fz}. The target is to testify the distributional dependences of redshif, luminosity and event rate of SGRBs on diverse SFR models. It can be seen from Figure \ref{fig:fz} that the derived redshift distributions of SFRs do have somewhat differences between two theoretical SFR models of \cite{2008ApJ...683L...5Y} and \cite{2014ARA&A..52..415M}.

\subsection{Luminosity function evolution of diverse time-delay models}
\begin{table*}[htbp]
    \centering
    \caption{The best fitting parameters of luminosity distributions }
    \renewcommand\arraystretch{1}
    \tabcolsep=0.2cm
    \begin{tabular}{c|c|c|c|c|c|c|c|c}
    \hline
    \hline
 &  Detector    &  Model      & $k_{1} $      & $k_{2} $      & $k_{3}  $     & $logL_{b1}  $   & $logL_{b2}  $      & $\chi^{2}/dof$        \\
  	\hline
\multirow{15}{*}{$\rho_{0,L}$} & \multirow{5}{*}{Swift} & Y{\"u}ksel+2008 & -1.85$\pm$0.07    & -1.44$\pm$0.17    & -1.31$\pm$0.21    & 50.11$\pm$0.39 & 51.19$\pm$1.26 & 2.17 \\
&       & Gauss  & -2.11$\pm$0.22    & -1.45$\pm$0.14    & -1.16$\pm$0.08    & 49.13$\pm$0.25 & 50.39$\pm$0.42 & 2.18 \\
&       & Lognormal    & -2.09$\pm$0.26    & -1.29$\pm$0.19    & -1.13$\pm$0.07    & 49.09$\pm$0.25 & 50.19$\pm$0.72  & 2.21 \\
&       & Power-law     & -1.90$\pm$0.26    & -1.48$\pm$0.11    & -1.27$\pm$0.09    & 49.03$\pm$0.40 & 50.58$\pm$0.59 & 2.13 \\
&       & Madau+2014     & -1.76$\pm$0.05    & -1.22$\pm$0.90    & -1.32$\pm$0.13    & 50.63$\pm$0.70 & 51.11$\pm$1.12 & 2.16 \\
\cline{2-9}
& \multirow{5}{*}{Fermi} & Y{\"u}ksel+2008 & -1.53$\pm$0.20    & -0.16$\pm$1.40    & -1.36$\pm$0.08    & 51.24$\pm$0.34 & 51.68$\pm$0.24 & 2.67 \\
&       & Gauss  & -1.25$\pm$0.18    & -0.17$\pm$0.41    & -1.29$\pm$0.09    & 51.21$\pm$0.21 & 51.85$\pm$0.13  & 2.55 \\
&       & Lognormal    & -1.10$\pm$0.18    & 0.08$\pm$0.50     & -1.27$\pm$0.08    & 51.19$\pm$0.21 & 51.78$\pm$0.12 & 2.56 \\
&       &Power-law     & -1.28$\pm$0.19    & -0.21$\pm$0.70    & -1.32$\pm$0.08    & 51.22$\pm$0.27 & 51.74$\pm$0.17 & 2.62  \\
&       & Madau+2014      & -1.52$\pm$0.19    & -0.09$\pm$2.54    & -1.34$\pm$0.08    & 51.29$\pm$0.57 & 51.67$\pm$0.25 & 2.65 \\
\cline{2-9}
& \multirow{5}{*}{Konus-wind}  & Y{\"u}ksel+2008 & -2.38$\pm$0.17   & -0.75$\pm$0.59    & -1.63$\pm$0.16    & 51.98$\pm$0.26 & 52.69$\pm$0.26 & 2.82 \\
&       & Gauss  & -2.29$\pm$0.14    & -0.52$\pm$0.35   & -1.47$\pm$0.14    & 51.94$\pm$0.17  & 52.73$\pm$0.17  & 2.59 \\
&       & Lognormal    & -2.16$\pm$0.14    & -0.31$\pm$0.35    & -1.47$\pm$0.14    & 51.94$\pm$0.16 & 52.73$\pm$0.14 & 2.60 \\
&       & Power-law      & -2.15$\pm$0.16    & -0.61$\pm$0.52    & -1.57$\pm$0.15    & 51.97$\pm$0.25 & 52.69$\pm$0.22 & 2.73 \\
&       &Madau+2014     & -2.27$\pm$0.15    & -0.81$\pm$0.73    & -1.59$\pm$0.15    & 52.01$\pm$0.30 & 52.66$\pm$0.32 & 2.76 \\
\hline
\multirow{15}{*}{$\rho_{0,>L}$   } & \multirow{5}{*}{Swift} & Y{\"u}ksel+2008 & -1.27$\pm$0.12    & -0.59$\pm$1.19    & -0.55$\pm$0.03    & 49.30$\pm$0.64 & 49.66$\pm$1.72 & 1.63 \\
&       & Gauss  & -1.12$\pm$0.05    & -0.35$\pm$0.01    & -0.75$\pm$0.43    & 49.27$\pm$0.04 & 51.66$\pm$0.47 & 1.93 \\
&       & Lognormal    & -0.91$\pm$0.05    & -0.29$\pm$0.01    & -0.75$\pm$0.25    & 49.20$\pm$0.04 & 51.57$\pm$0.24 & 1.94   \\
&       & Power-law     & -0.99$\pm$0.04    & -0.43$\pm$0.01    & -0.76$\pm$3.29    & 49.31$\pm$0.05 & 51.76$\pm$3.99 & 1.92 \\
&       & Madau+2014      & -1.17$\pm$0.06    & -0.67$\pm$0.06    & -0.48$\pm$0.03   & 49.32$\pm$0.11 & 50.23$\pm$0.19 & 1.91 \\
\cline{2-9}
& \multirow{5}{*}{Fermi} & Y{\"u}ksel+2008 & -0.48$\pm$0.06  & -0.29$\pm$0.02 & -0.87$\pm$0.14 & 50.98$\pm$0.20 & 52.75$\pm$0.14 & 2.17  \\
&       &Gauss  & -0.19$\pm$0.01 & -0.34$\pm$0.09 & -0.86$\pm$0.14 & 52.09$\pm$0.34 & 52.82$\pm$0.14 & 2.13 \\
&       & Lognormal    & -0.14$\pm$0.03 & -0.41$\pm$0.06 & -1.02$\pm$0.21 & 52.08$\pm$0.16 & 53.01$\pm$0.15 & 2.09 \\
&       & Power-law     & -0.26$\pm$0.01 & -0.76$\pm$0.37 & -1.01$\pm$0.63 & 52.71$\pm$0.28  & 53.17$\pm$1.09 & 2.21 \\
&       & Madau+2014      & -0.50$\pm$0.07    & -0.30$\pm$0.02    & -0.98$\pm$0.20    & 51.02$\pm$0.23 & 52.88$\pm$0.14 & 2.23 \\
\cline{2-9}
& \multirow{5}{*}{Konus-wind}  & Y{\"u}ksel+2008 & -1.25$\pm$0.16 & -0.41$\pm$0.07 & -0.73$\pm$0.09 & 51.36$\pm$0.12 & 52.73$\pm$0.27 & 2.65 \\
&       & Gauss  & -1.03$\pm$0.09 & -0.24$\pm$0.04 & -0.64$\pm$0.06 & 51.29$\pm$0.07 & 52.73$\pm$0.15 & 2.36 \\
&       & Lognormal    & -0.80$\pm$0.08 & -0.19$\pm$0.03 & -0.64$\pm$0.06 & 51.23$\pm$0.08  & 52.73$\pm$0.12 & 2.38  \\
&       &Power-law     & -0.94$\pm$0.11 & -0.32$\pm$0.05 & -0.69$\pm$0.07 & 51.28$\pm$0.11  & 52.73$\pm$0.19 & 2.47\\
&       & Madau+2014      & -1.20$\pm$0.16    & -0.44$\pm$0.07    & -0.70$\pm$0.09    & 51.33$\pm$0.14 & 52.73$\pm$0.34 & 2.72 \\
\hline
    \end{tabular}%
    \label{tab:fitting}%
\end{table*}%

\begin{figure}
	\centering
	\includegraphics[width=\textwidth]{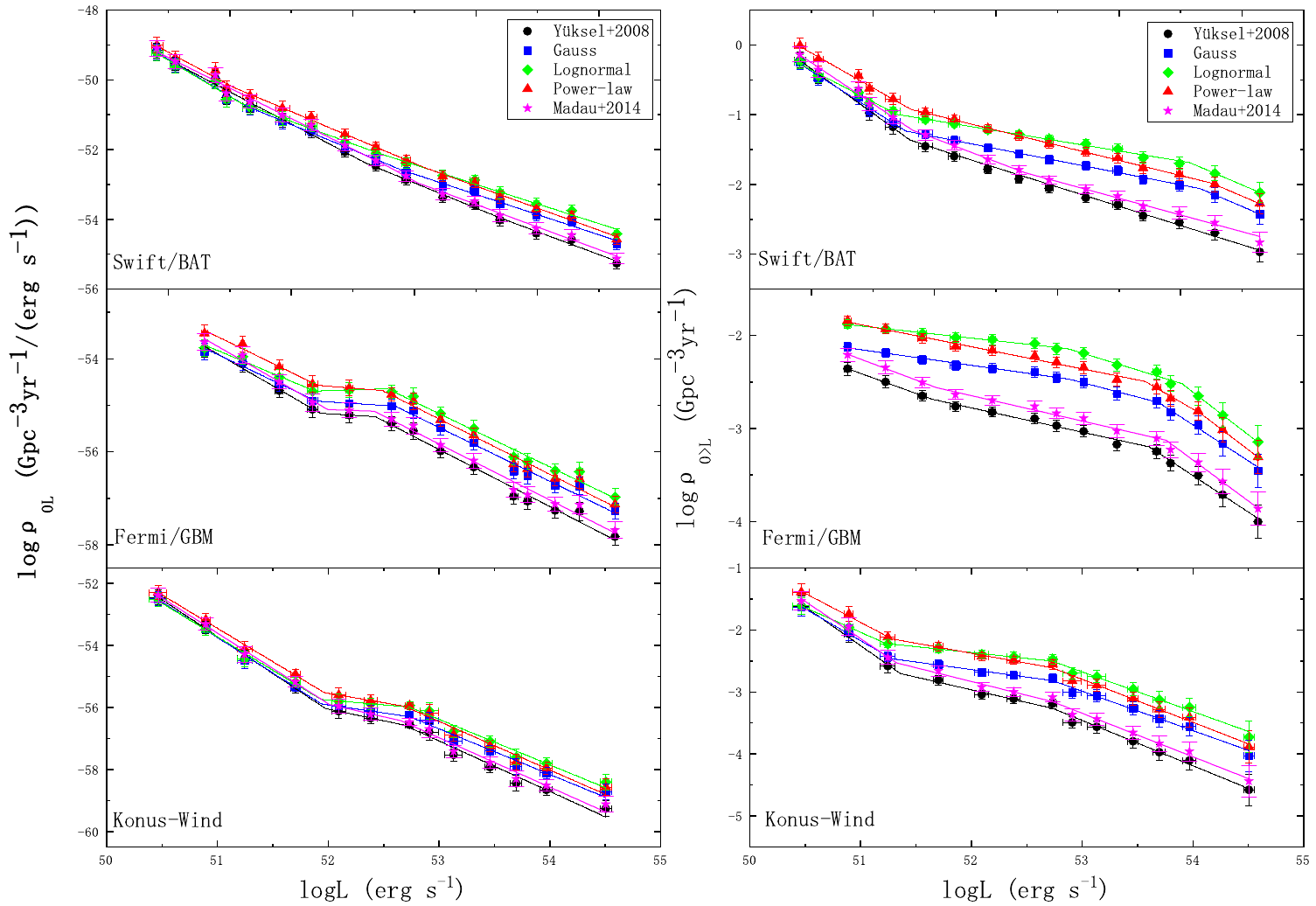}
	\caption{The logarithmic relations between the local event rate density and the luminosity of SGRB samples detected by Swift (upper panels), Fermi (middle panels) and Konus-wind (lower panels) satellites. Two undelayed SFR models of \cite{2008ApJ...683L...5Y} and \cite{2014ARA&A..52..415M} and three delayed SFR models built on \cite{2008ApJ...683L...5Y} for  Gauss, Lognormal and Power-law cases are symbolized by filled circles, stars, squares, diamonds and triangles, respectively. The broken lines stand for the best fits with a triple power-law form of Eq. (\ref{rou-L}).} \label{fig:3}
\end{figure}

Now we investigate how the local event rate density evolves the luminosity for each GRB sample based on the redshift distributions provided in Eqs. (\ref{fz1}-\ref{fz3}). It is shown in Figure \ref{fig:3} that the local event rate densities of SGRBs detected by different satellites decrease with the increase of luminosities. The left panels depict the relation between the luminosity and the local event rate density at a specific luminosity. The right panels show the relationships between the luminosity and the local event rate density above a given luminosity. Both relations can be fitted by a triple power-law function as
\begin{equation} \label{rou-L}
	log\rho_{0}=\left\{
	\begin{array}{rcl}
&A_1+k_1(logL-logL_{b1}),& [{L\leqslant L_{b1}}]\\
&A_1+k_2(logL-logL_{b1}),& [{L_{b1}<L<L_{b2}}]\\
&A_1+k_3log(L/L_{b2})+k_2log(L_{b2}/L_{b1}),& [{L\geqslant L_{b2}}].\\
	\end{array} \right.
\end{equation}
The best fitting results are listed in Table \ref{tab:fitting}, where we notice that the distributions of local event rates vary from diverse satellites because they are dependent of energy bands. By comparison, Swift SGRBs hold relatively smaller luminosites than Fermi/Konus-wind SGRBs. It needs to point out that we have only taken into account the luminosity errors and Poisson errors of the local event rate density, so the actual errors will be larger and the actual chi-squares will be smaller than the current ones. It is also found that the event rates estimated by five theoretical models exhibit the similar trend in each panel. In contrast, the event rate densites derived with the delayed models are slightly larger especially for those SGRBs with higher luminosities. It is worthy of attention that two undelayed SFR models evolve with luminosity almost in the same manner.

\subsection{Local event rate densities of different SGRB samples}
For the above SGRB samples, we apply Eq. (\ref{R}) in \cite{2015ApJ...812...33S} to estimate the local event rates within distinct time-delay frameworks and list the results in Table \ref{tab:local}. It is necessary to point out that the derived local rates should be the lower limits once the instrument downtime of a detector is reckoned in when the satellite was in South Atlantic Anomaly (SAA) region. Considering the beaming factor of $f_B=1-cos\theta_j$ for a relativistic jet with half-opening angle $\theta_j$, the actual local event rate should be $\rho_{0,ture}=\rho_0/f_B$. Assuming a typical value of   $\theta_j\approx9.8\pm1.3$ degrees \citep{2006ApJ...650..261S,2018PASP..130e4202Z}, we estimate the jet-corrected event rate densities to be 0.30-66.47 $\rm Gpc^{-3} \rm yr^{-1}$ once the variety of local event rate densities for diverse time delay models in Table \ref{tab:local} are taken into account. Note that our event rates accord with some previous results but span a relatively larger range than $ 4-7 \; \mathrm{Gpc^{-3} yr^{-1}}$ \citep{2015MNRAS.448.3026W,2015ApJ...812...33S} or $ 15 \; \mathrm{Gpc^{-3} yr^{-1}}$ \citep{2020ApJ...896...83G} for Swift SGRBs, $ 40 \; \mathrm{Gpc^{-3} yr^{-1}}$ for CGRO/BATSE SGRBs  \citep{2006ApJ...650..281N}, $7.53\; \mathrm{Gpc^{-3} yr^{-1}}$  \citep{2018ApJ...852....1Z} or  $17.43\; \mathrm{Gpc^{-3} yr^{-1}}$ \citep{2021RAA....21..254L} for Fermi/GBM SGRBs. In particular, we notice that the merger rate of $35\; \mathrm{Gpc^{-3} yr^{-1}}$ for the BH-NS binaries \citep{2021ApJ...917...24Z} falls into our range of SGRB rates.  

There are several factors that may evidently affect the accuracy of the local event rate estimation. Firstly, selection effects including sample selection standards, luminosity function selection and different SFR models might play an important role. Secondly, instrument effects comprising energy bands, instrument uptime and threshold of detectors will results in larger diversity of GRB rate estimate. Thirdly, the off-axis effect will influence the estimate of GRB rates especially for those low-luminosity GRBs viewed sideways. In addition, the deduced event rates are could be also biased by different methods or detectors more or less. The relatively larger range of our local event rates are mainly attributed to detetors with diverse energy bands and sensitivities.

\begin{table*}[htbp]
	\centering
	\setlength{\belowcaptionskip}{0cm}
	\caption{The local event rate density of SGRBs detected by diverse detectors }
	\scriptsize
	\begin{tabular}{c|c|c|c|c|c|c}
		\hline
		\hline
		\multirow{2}{*}{Detector}&\multicolumn{5}{c|}{Local event rate $\rho_{0}$ in unit of $\rm Gpc^{-3} \rm yr^{-1}$}&$L_{min}$\\ \cline{2-6} 
	 &Y{\"u}ksel+2008& Gauss-Y{\"u}ksel  & Lognormal-Y{\"u}ksel & Power-law-Y{\"u}ksel &Madau+2014&  ($\rm erg\ \rm s^{-1}$)\\
		\hline
		Swift/BAT &  0.60 $\pm$ 0.18   & 0.56 $\pm$ 0.16  & 0.57 $\pm$ 0.15&0.97 $\pm$ 0.26 & 0.73 $\pm$ 0.18 & 2.57$\times 10^{48}$\\
		Fermi/GBM      & $(4.35\pm0.74)\times 10^{-3}$  & $(7.56\pm0.90)\times 10^{-3}$  &$(1.31\pm0.14)\times 10^{-2}$ &$(1.43\pm0.19)\times 10^{-2}$ & $(6.18\pm0.98)\times 10^{-3}$   & 1.97$\times 10^{50}$\\
		Konus-Wind & $(2.41\pm1.04)\times 10^{-2}$   & $(2.30\pm0.94)\times 10^{-2}$   & $(2.45\pm0.88)\times 10^{-2}$  & $(4.02\pm1.51)\times 10^{-2}$&   $(2.92\pm0.78)\times 10^{-2}$ &2.91$\times 10^{50}$\\
		\hline
		\hline 
	\end{tabular}%
	\label{tab:local}%
\end{table*}%


\subsection{The SGRB rates versus the delayed/undelayed SFRs}
\begin{figure}
	\centering
	\includegraphics[width=0.46\textwidth]{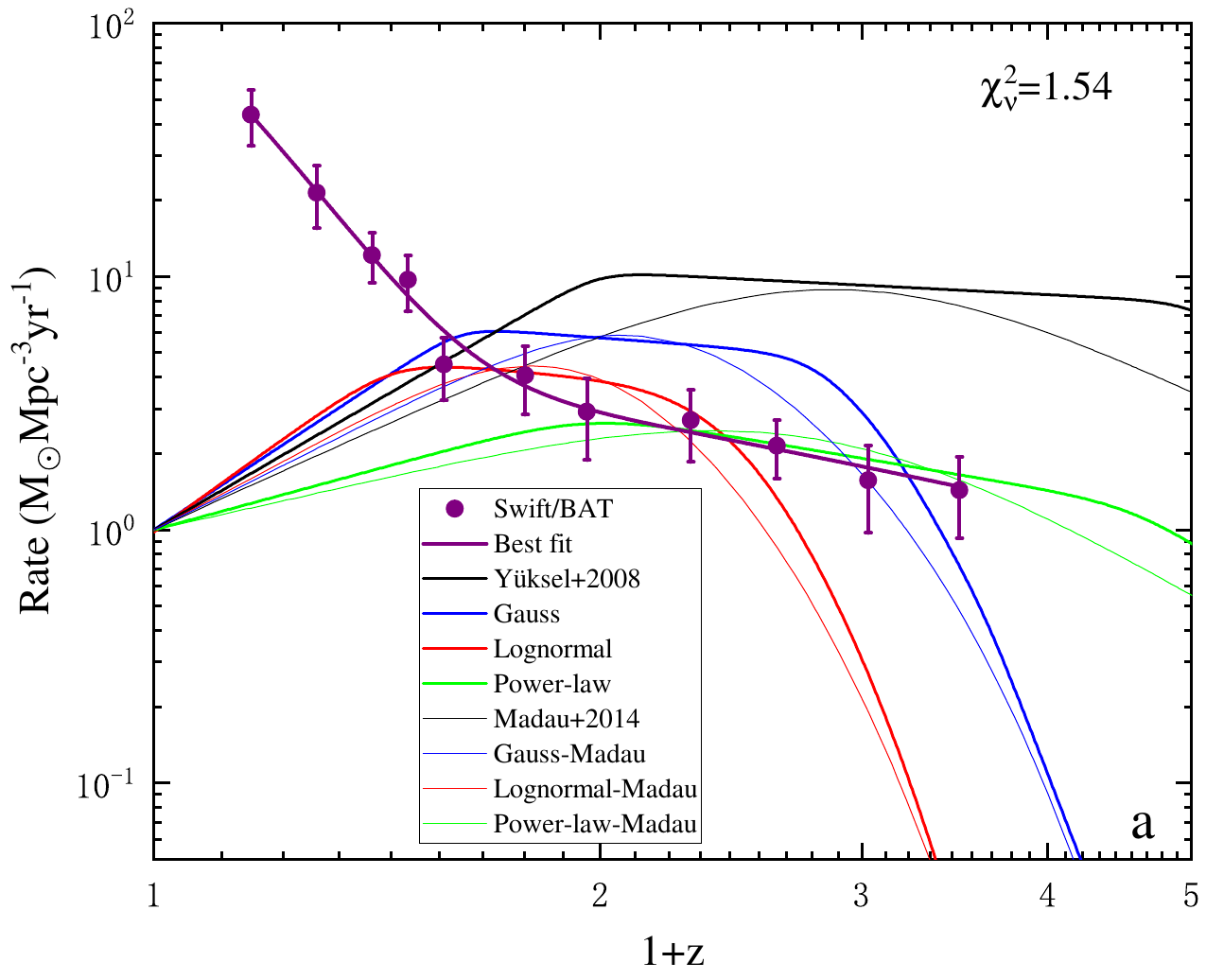}
	\includegraphics[width=0.46\textwidth]{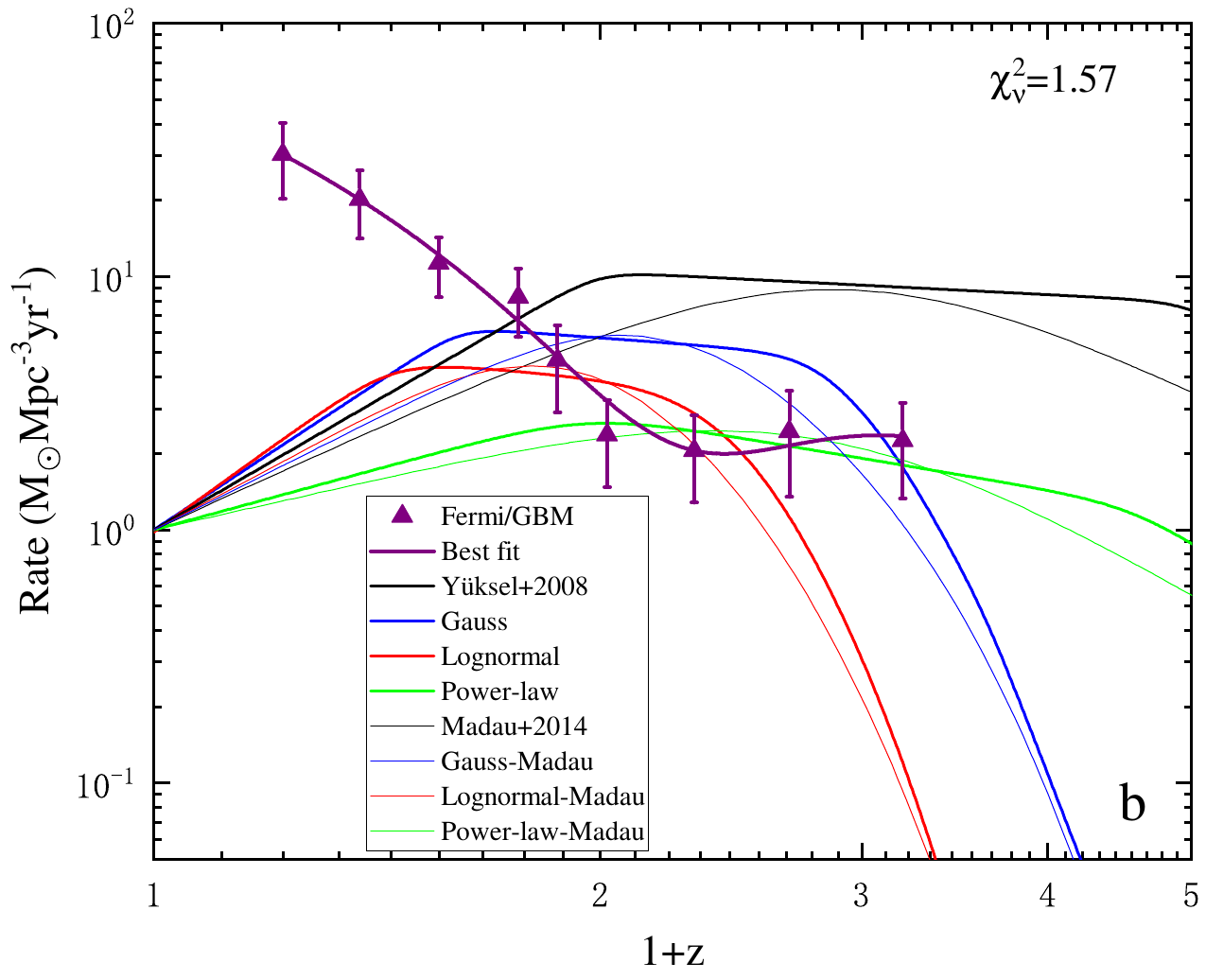}
	\includegraphics[width=0.46\textwidth]{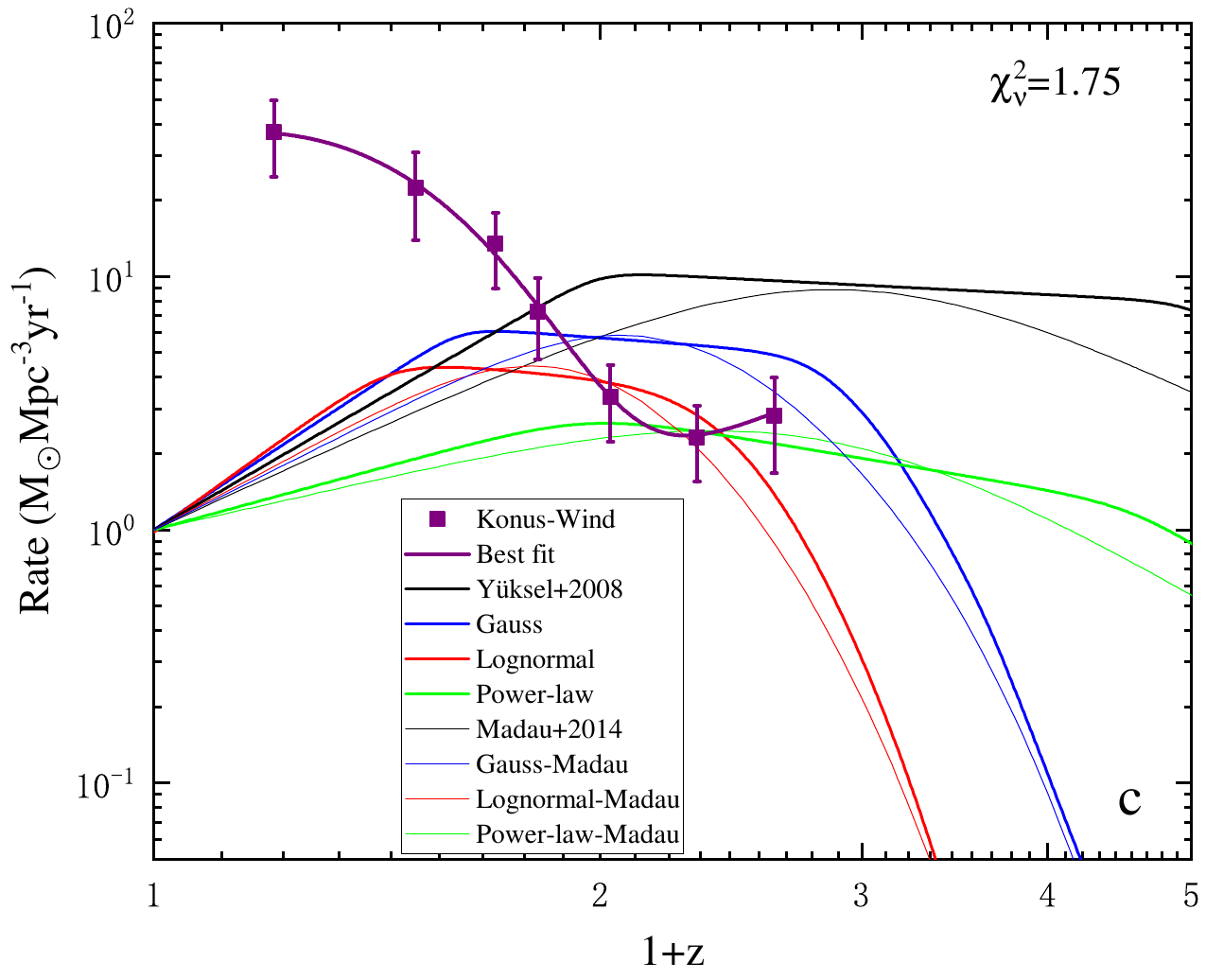}
	\caption{Comparison of the observed SGRB rates (filled squares) of Swift (Panel a), Fermi (Panel b) and Konus-wind (Panel c) satellites with different kinds of SFRs as shown in Figure \ref{fig:fz}. The thick lines depict the SFR models related with  \cite{2008ApJ...683L...5Y} and the thin lines indicate the SFR models derived from \cite{2014ARA&A..52..415M}. The purple solid curves represent the best fits to the SGRB rates with Eq. (\ref{double}). \label{fig:R}}
\end{figure}

\begin{figure}
	\includegraphics[width=\textwidth]{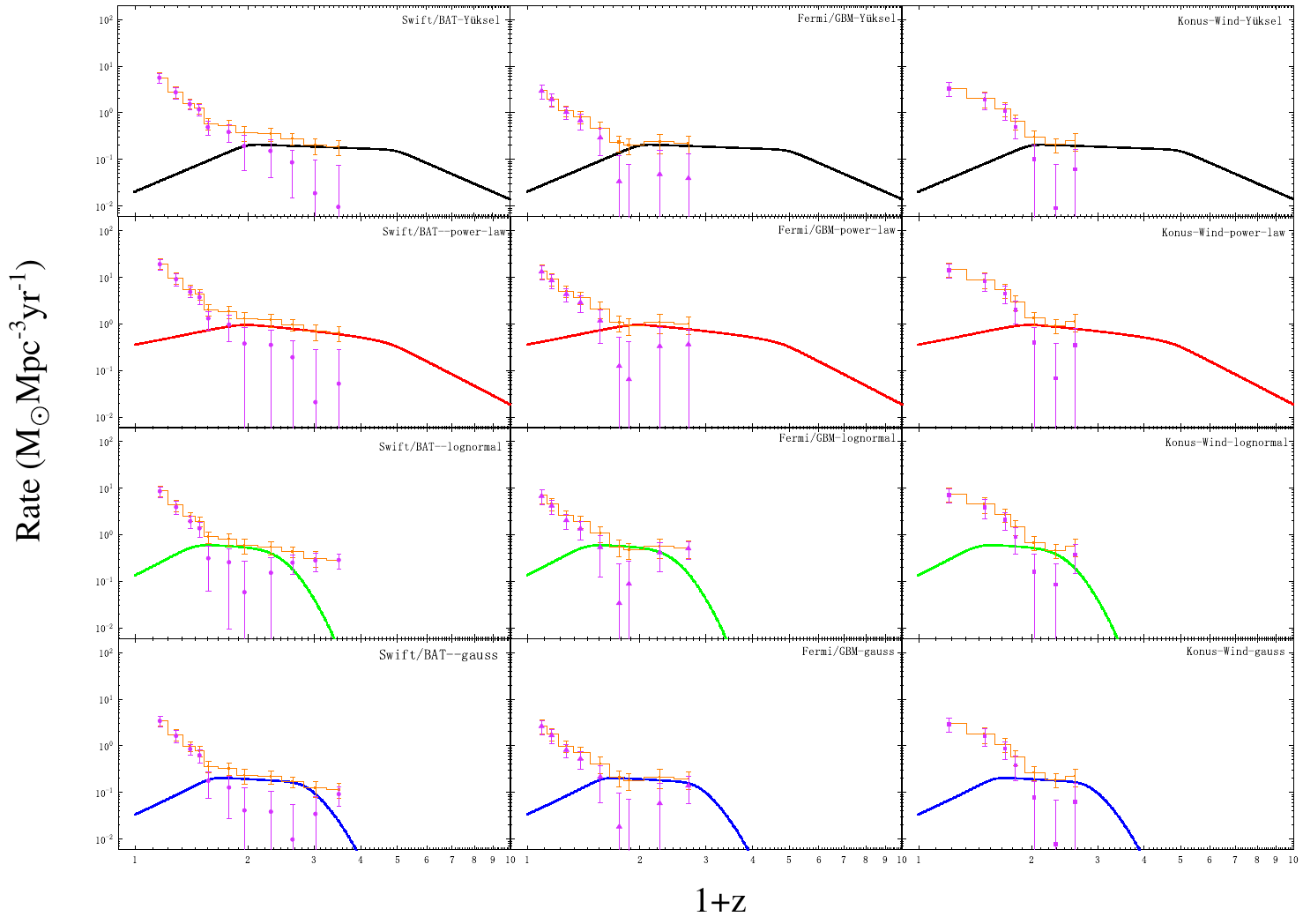}
	\caption{Comparison of the SGRB event rates (step lines) reported by Swift/BAT (left Panels), Fermi/GBM (middle Panels) and Konus-wind (right Panels) detectors with the undelayed SFR of \cite{2008ApJ...683L...5Y} on the first line and the delayed SFRs in downward order for the Power-law, Lognormal and Gaussian delay models, respectively. The purple squares represent the extra components when the SFRs are subtracted from the observed SGRB event rates. \label{fig:ALL}}
\end{figure}

Combining the distributions of redshift and luminosity, we calculate the event rates of three SGRB samples with Eq. (\ref{R}) and compare them with different time-delayed SFR models in Figure \ref{fig:R}, from which we find that the SGRB event rates of diverse detectors are roughly consistent with the SFRs with/without time-delays at higher redshifts and significantly exceed the the SFRs at a redshift lower than $\sim1$. In terms of the trend of rate evolution with redshift, the lognormal delay model of both \cite{2008ApJ...683L...5Y} and \cite{2014ARA&A..52..415M} can not be solely adopted to describe the observations even at the higher redshift end. We now fit the observed event rates with a two-component function below
 
\begin{equation}\label{double}
	\begin{aligned}
	R_{SGRB}(Z)=\frac{A_1}{C\sqrt{2\pi} } \exp \left[-\frac{(Z-B)^{2}}{2C^{2} }\right]+A_2 Z^{-D},
	\end{aligned}
\end{equation}
where $A_1$, $A_2$, $B$, $C$ and $D$ are the fitted parameters and $Z=1+z$. The least Chi-square values have been put on each panel of Figure \ref{fig:R}. Our results support the finding by \cite{2021ApJ...914L..40D} that the SGRBs rate exceeds the power-law delayed SFR at lower redshifts. Furthermore, we also find that the Lognormal and Gaussian delayed models and some undelayed SFR models  \citep[e.g.][]{2008ApJ...683L...5Y,2014ARA&A..52..415M} can not be fully excluded despite of the slghtly larger deviations with respect to the observed SGRB rates. One can qualitatively compare the SGRB event rate with the SFR model on the assumption that the observed event rate is uniquely connected with any one of the SFR models. As displayed in Figure \ref{fig:ALL}, we notice that both the undelayed SFR of \cite{2008ApJ...683L...5Y} together with its diversely delayed SFRs including the Power-law and Gaussian models, except the Lognormal delay model, match the observed event rate well. However, neither Lognrmal nor Gaussian delay SFR model of \cite{2014ARA&A..52..415M} identifies with the observed SGRB rate within all redshift ranges. This demonstrates that the SGRB event rates can be interpreted by not only the time-delayed SFR models but also the undelayed SFR templates, manifesting the complexity and miscibility of SGRB progenitors. Simultaneously, a power-law-like decay event rate emerges in each case when the corresponding SFR parts are removed. The extra components at lower redshifts obviously exceed the SFRs, which can be contributed by the older star populations such as the compact binary mergers or those low-luminosity sources. Interestingly, this is analogous to the event rate of non-repeating Fast Radio Bursts (FRBs) found by \cite{2024arXiv240600476Z}. These findings enable us to speculate that SGRBs and FRBs could be associated with each other physically.


\begin{figure}
	\includegraphics[width=\textwidth]{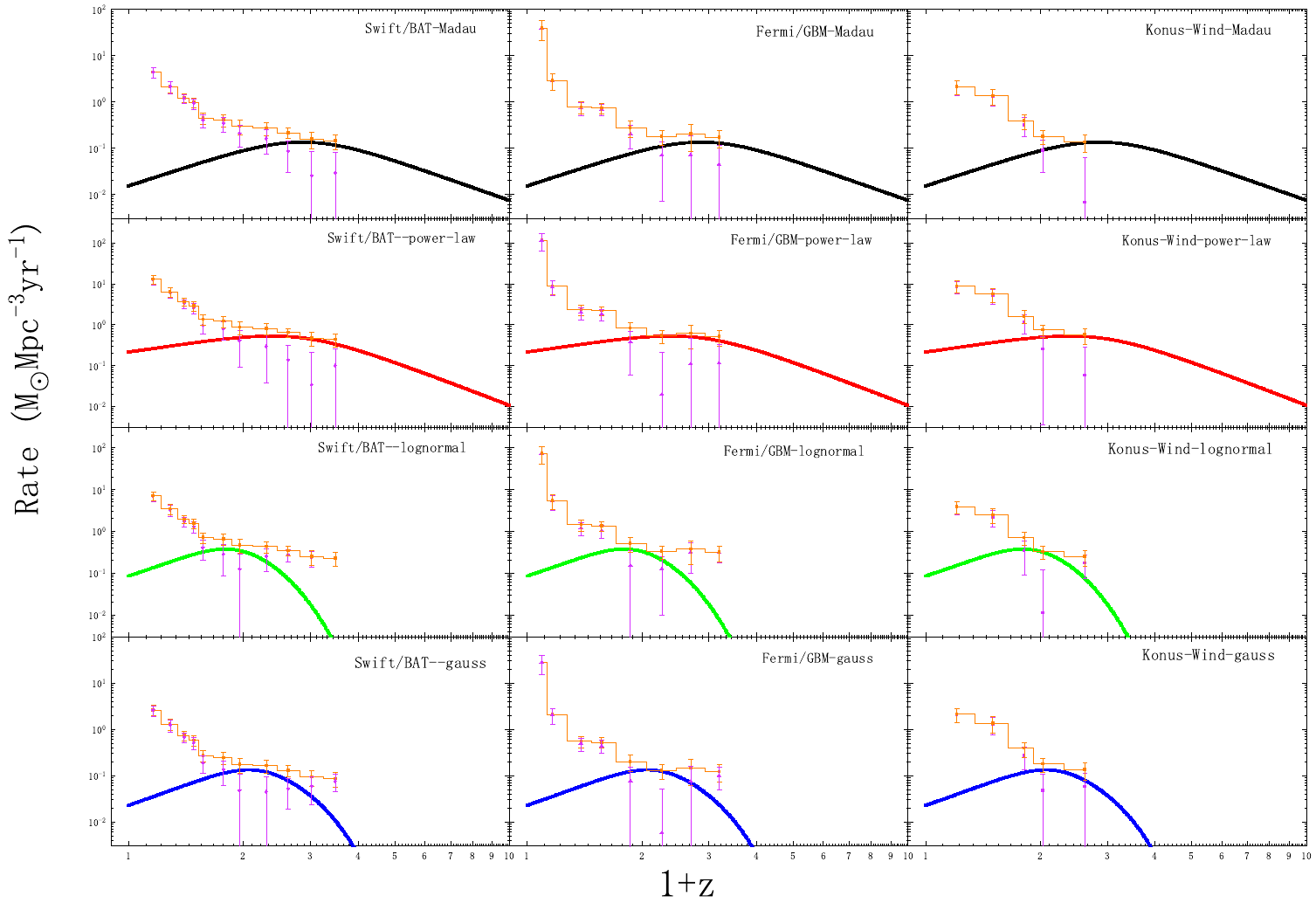}
	\caption{Comparison of the SGRB event rates (step lines) reported by Swift/BAT (left Panels), Fermi/GBM (middle Panels) and Konus-wind (right Panels) detectors with the undelayed SFR of \cite{2014ARA&A..52..415M} on the first line and the delayed SFRs in downward order for the Power-law, Lognormal and Gaussian delay models, respectively. All symbols are the same as in Figure \ref{fig:ALL}.\label{fig:ALL-madau}}
\end{figure}
\section{Conclusion and discussions} \label{sec:five}
In this paper, we have carefully investigated the effect of sample selection on the luminosity functions and event rates of SGRBs and compared the GRB event rates with the distinct SFR histories in a robust way. We can draw the following conclusions:
\begin{itemize}
	\item Observationally, we find that the redshift and luminosity distributions of Fermi/GBM and Konus-wind SGRBs are identical and they are different from those of Swift/BAT SGRBs, indicating the redshift and luminosity distributions rely on the energy bands of detector in evidence. The fractions of SGRBs with lower redshift and/or lower luminosity are relatively high.
	\item The luminosity distributions of SGRBs detected by diverse detectors can be fitted by a smoothly broken power-law function. The broken luminosities are around $2.14 \times 10^{50} \mathrm{erg\; s^{-1}}$, $6.46 \times 10^{51} \mathrm{erg\; s^{-1}}$ and $5.43 \times 10^{52} \mathrm{erg\; s^{-1}}$ for Swift/BAT, Fermi/GBM and Konus-wind SGRBs, correspondingly. The median luminosity of Swift SGRBs is about one order of magnitude smaller than that of Fermi/GBM or Konus-wind SGRBs. 
	\item We find that the local event rates of Swift/BAT, Fermi/GBM and Konus-wind SGRBs are around two orders of magnitude larger than that of either Fermi or Konus-wind SGRBs, while the latter two event rates approach with each other in spite of more scattering of Fermi SGRBs across different SFR models. Our estimats are basically consistent with previous results in a wider parameter space.
	\item The relations of the observed SGRB rate versus redshift are successfully fitted by a two-component model of power-law plus Gauss function. We find that the rate evolution of SGRBs of three kinds of detectors matches the delayed/undelayed SFRs well except the delayed Lognormal and Gaussian SFRs of \cite{2014ARA&A..52..415M} or the only delayed Lognormal SFR of \cite{2008ApJ...683L...5Y} at higher redshifts and exceeds all types of SFRs at lower redshits. 
	\item To check the inconsistency of the observed SGRB rates with the SFRs quantitatively, we deduct the diverse SFR components from the SGRB rates and find that the remaining SGRB rates steeply decline with redshift in a power-law-like form. It is interestingy found that the observed SGRB rates exceed not only the undelayed SFRs but also the delayed SFRs in the lower redshift range of $z<1$ except that the Lognormal and/or Gaussian delay forms seem incoherent with the observations. 
\end{itemize}

The low-redshift excess of event rate with respect to the SFR has been found in many objects including LGRBs \citep[e.g.][Liu et al. 2025]{2015ApJS..218...13Y,2015ApJ...806...44P,2022MNRAS.513.1078D}, SGRBs \cite[][this work]{2018ApJ...852....1Z}, fast radio bursts (FRBs, \cite[e.g.][Pan et al. 2025]{2024arXiv240600476Z}), black holes (BH, \cite{2024ApJ...977...29D}) and Active Galactic Nuclei (AGN, Rong et al. 2025). These objects are largely different in size from stellar to galactic distance scales. An important issue is how to explain the low-redshift excess physically. \cite{2023ApJ...958...37D} verified that the discrepancy is attributed to the diversity of luminosities, in other words, the event rates of low-luminosity LGRBs exceed the SFR while the high-luminosity LGRBs match the star formation history excellently no matter whether the LGRB samples are complete or not. Later, \cite{2024ApJ...963L..12P} found that  part of LGRBs with higher event rate have a tendency aligning with a delay SFR, which enables them to supporse that the low-redshift LGRBs may have originated from a compact binary merger. It should be emphasized that the low-luminosity LGRBs may have larger redshifts and those high-luminosity LGRBs might have smaller redshifts although most low-luminosity LGRBs have the lower redshifts of which the majority of SGRBs possess. However, it can be seen in this work that the event rates of SGRBs can be characterized by both the undelayed SFR model and the distinctly delayed SFR models, suggesting that a fraction of SGRBs might be formed by the core-collapse of massive stars rather than the compact bianary mergers \cite[see also][]{2022ApJ...940....5D}. This strongly demonstrates that the progenitors of SGRBs could be diverse and more complicated than the dichotomy of collapse and non-collapse.

In the aspect of higher event rate of SGRBs, we suggest that their higher rates at lower redshifts can be interpreted by the off-axis effect or the larger half-opening jet angles owing to their impirical energy correlations with larger scaters (e.g. Zhang et al. 2025). Moreover, the possibility of higher event rats at lower redshifts due to the instrumental effect or the flux limit can not be excluded in that some SGRBs lying in remote distance will be absent from our observed samples. In this situation, a volume-limited sample will be of benefit to the event rate estimation as shown in Zhang et al. (2025). Of course, the jet structure of a SGRB will also act on the event rate more or less. Unfortunately, the off-axis angles have been confirmed only for two GRBs nowadays, namely the kilonova-associated SGRB 170817A/GW 170817 from a binary neutron star (BNS) merger accompanying with a structured jet \citep{2017Natur.551...85A,2018ApJ...867..147W,2017PhRvL.119p1101A,2018Natur.561..355M} and the Supernova-associated GRB 171205A \citep{2024ApJ...962..117L}.

Although part of SGRBs can be characterized by any one of these SFR models, it is very hard to judge which model is the real one. Different SGRBs are likely explained by diverse SFR models or their mixture. This can be supported by some peculiar GRBs, such as SGRBs with collapsar origin but long duration, LGRBs from merges with short duration, etc.   It is generally believed that LGRBs originate from the core-collapse of massive stars, while SGRBs are produced by the merger of compact stars. However, the parameter distributions of two kinds of GRBs are usually overlapped, which causes it difficult to distinguish between the undelayed SFRs for collapsars and the delayed SFRs related with non-collapsars. Even for the non-collapsing events, people still have difficulties of identifying diverse compact bianary mergers. For example, \cite{2022RAA....22g5011Y} proposed that GRB 201221D was generated from a BNS merger, while others believed that this burst originated from a BH–NS merger \citep{2022MNRAS.514.2716M}. GRB 130603B as another typical burst had given rise to many studies of its progenirtors involving BNS mergers \citep{2014MNRAS.439.3916M}, BH-NS mergers \citep{2016ApJ...825...52K}, magnetar \citep{2023ApJ...949L..32D}, and so on. Therefore, the roadmap to reveal the relation of the SGRB rate with the SFR models should rely on the joint observations in multi-wavelengths and multi-missions in the future.

\section{Acknowledgements}
We are very grateful to the anonymous referee for valuable comments and suggestions. This work was supported in part by National Natural Science Foundation of China
(grant No.U2031118). We acknowledge the usage of the archive data provided by Swift, Fermi and Konus-wind satellites.

\end{document}